\documentclass[%
 reprint,
superscriptaddress,
nofootinbib,
 amsmath,amssymb,
 aps,
 pra, longbibliography,
]{revtex4-2}
\usepackage{layouts}
\usepackage{graphicx, subfigure}
\usepackage{dcolumn}
\usepackage{bm}
\usepackage{mathtools}
\usepackage{amsmath}
\usepackage{pgfplots}
\usepackage{float}
\usepgfplotslibrary{fillbetween}
\pgfplotsset{compat=1.16}
\usetikzlibrary{fit,backgrounds,positioning,shapes,shapes.callouts}
    \usepackage[scaled=0.83]{helvet}
    \usepackage{marvosym,pifont}
    \usepackage[T1]{fontenc}
    \usepackage[utf8]{inputenc}

\usepackage{hyperref}

\begin{document}


\title{Theory of coherent interaction-free detection of pulses}

\author{John J. McCord}
\email[]{john.mccord@aalto.fi}

\affiliation{QTF  Centre  of  Excellence, 
 Department of Applied Physics, Aalto University, FI-00076 Aalto, Finland\\}

\author{Shruti Dogra}

\affiliation{QTF  Centre  of  Excellence, 
	Department of Applied Physics, Aalto University, FI-00076 Aalto, Finland\\}

\author{Gheorghe Sorin Paraoanu}

\affiliation{QTF  Centre  of  Excellence, 
	Department of Applied Physics, Aalto University, FI-00076 Aalto, Finland\\}

\date{\today}

\begin{abstract}

Quantum physics allows an object to be detected even in the absence of photon absorption by the use of so-called interaction-free measurements. We provide a formulation of this protocol using a three-level system, where the object to be detected is a pulse coupled resonantly to the second transition. In the original formulation of interaction-free measurements, the absorption is associated with a projection operator onto the third state. We perform an in-depth analytical and numerical analysis of the coherent protocol, where coherent interaction between the object and the detector replaces the projective operators, resulting in higher detection efficiencies. We provide approximate asymptotic analytical results to support this finding. We find that our protocol reaches the Heisenberg limit when evaluating the Fisher information at small strengths of the pulses we aim to detect -- in contrast to the projective protocol that can only reach the standard quantum limit. We also demonstrate that the coherent protocol remains remarkably robust under errors such as pulse-rotation phases and strengths, the effects of relaxation rates and detunings, as well as different thermalized initial states. 

\end{abstract}

\pacs{Valid PACS appear here}
\maketitle

\section{Introduction}

Interaction-free measurements  \cite{Elitzur_1993} are a type of quantum hypothesis test where the presence of an object is confirmed or falsified even when the probe photons are not absorbed. As originally formulated, interaction-free measurements are based on the
observation that placing an ultrasensitive object in one arm of a Mach-Zehnder interferometer alters the output probabilities, thus allowing us to probabilistically infer its presence. This class of measurements provides a remarkable illustration of so-called negative-result measurements, as first described by Renninger \cite{Renninger} and Dicke \cite{Dicke}.  Furthermore, the detection efficiency can be improved by utilizing the quantum Zeno effect \cite{peres-ajp-1980}  through repeated ``interrogations'' of the object \cite{Kwiat_1995, Kwiat_1999, Ma2014, Peise2015}. 

Several topics in the foundations of quantum mechanics have been motivated by interaction-free measurements, such as Hardy's paradox \cite{Hardy_1992} -- where they have been utilized to rule out local hidden variables. Others include developments in quantum thermodynamics \cite{Elouard_2020} -- where an engine is proposed that is able to do useful work on an Elitzur-Vaidman bomb without seemingly having interacted with it. Finally, interaction-free measurements can induce non-local effects between distant atoms \cite{Aharonov_2018}, while
the Zeno effect has been shown to transform a single qubit gate operation into multi-qubit entangling gates, even in non-interacting systems \cite{Blumenthal_2022, Burgarth_2014}. Various implementations of interaction-free measurements have been done on different experimental platforms, leading to a plethora of applications. Some examples include optical imaging, where a photosensitive object is imaged in an interaction-free manner \cite{White1998}; counterfactual quantum cryptography, where a secret key distribution can be acquired without any particle carrying this information being transmitted through a quantum channel \cite{Noh2009, zheng-pra-2020}; and counterfactual ghost-imaging --  where ghost imaging, i.e., the technique of using entangled photon pairs for detecting an opaque object with improved signal-to-noise ratio, is merged with the idea of interaction-free measurements. This combined technique significantly reduces photon illumination and maintains comparable image quality of regular ghost imaging \cite{Zhang_2019, hans-npj-2021}. A related idea -- combining interaction-free measurements with the concept of induced coherence -- led to the realization of single-pixel quantum imaging of a structured object with undetected photons \cite{Ma2023}. Other examples include counterfactual communication \cite{salih-prl-2013,Vaidman_2015, Cao2017,vaidman-pra-2019, Walther2019, Aharonov2021} and counterfactual quantum computation \cite{Hosten2006}. Remarkably, these advancements have shown that information can be transmitted independent of a physical particle carrying it \cite{salih-prl-2013, Cao2017, Walther2019}. Overall, these results demonstrate that the interaction-free concept offers an unconventional yet viable avenue towards quantum advantage: tasks that manifestly cannot be achieved classically can be realized in this framework.

We recently proposed and experimentally demonstrated a novel protocol \cite{our_protocol} which employs repeated coherent interrogations instead of projective ones as used in the original interaction-free concept \cite{Elitzur_1993,peres-ajp-1980,Kwiat_1995, Kwiat_1999, Ma2014, Peise2015}. This distinction is fundamental and for clarity we will refer to the original protocol as ``projective'' and to ours as ``coherent''. We formulate these protocols as the task of detecting the presence of a microwave pulse in a transmission line via a resonantly-activated detector realized as a three-level (qutrit) transmon. We hereafter refer to these pulses as $B$ pulses, which are taken close to resonance with respect to the second transiton. 
The connection to a specific  superconducting-circuit implementation is convenient but not restrictive: indeed the concepts are general and can be readily employed in any experimental platform where a three-level system is available.
We investigate the theoretical foundations of this protocol, providing approximate analytical results in the asymptotic limit. We then study the sensitivity of the success probability and efficiency of our coherent protocol and compare with the corresponding figures of merit of the projective protocol under a variety of realistic experimental scenarios. Our results show that coherence acts as an additional quantum resource, allowing the accumulation of information about the $B$ pulses under successive exposures separated by Ramsey pulses on the first transition. Moreover, the proposed protocol can be further generalized to the detection of quantized $B$ pulses (such as photons in single or multiple cavities).

The paper is organized as follows: In Sec.~\ref{sec_2}, we introduce our coherent detection scheme and compare it with the standard projective scheme as often described in optical systems. We outline the two hypotheses: the system evolution with only beam-splitters and the evolution in the presence of pulses we wish to detect.  In Sec.~\ref{sec_3}, we investigate the limit when the number of Ramsey sequences $N$ is large, and subsequently explore the lower limit of $B$-pulse strength $\theta$ leading to sufficiently high detection efficiency. In Sec.~\ref{sec_4a}, we investigate how information on the presence of the pulses is acquired during each protocol by studying the success probabilities obtained when $B$ pulses of the same strength are applied. We also investigate, in Sec.~\ref{sec_4b}, the successive probabilities of successful interaction-free detection, as well as those of absorption, for $N = 25$ Ramsey sequences subjected to $B$ pulses of strength $\theta = \pi$. Additionally, in Sec.~\ref{sec_4c}, we investigate the quantum limits of each protocol by studying the Fisher information and the Fisher information of the efficiencies. In Sec.~\ref{sec_5}, we consider several sources of error and expound on their implications for the effectiveness of our protocol. These include the effect of beam-splitter strength (Sec.~\ref{sec_5a}), $B$ pulses with a variable phase  (Sec.~\ref{sec_5b}), interaction-free detection with randomly placed $B$ pulses (Sec.~\ref{sec_5c}), initialization in thermal states (Sec.~\ref{sec_5d}), effects of decoherence (Sec.~\ref{sec_5e}), and detuned $B$ pulses (Sec.~\ref{sec_5f}). 

\section{Coherent interaction-free measurements with qutrits} \label{sec_2}

Our protocol employs repeated coherent interrogations to perform interaction-free measurements using a qutrit \cite{our_protocol}. We consider a qutrit (three-level quantum system) with basis states 
$(|0\rangle, |1\rangle, |2\rangle)$ and introduce the asymmetric Gell-Mann generators of SU(3) by $\sigma^{y}_{kl} = -i|k\rangle \langle l| + i|l\rangle \langle k|$, $\sigma^{x}_{kl}= |k \rangle \langle l| + |l\rangle \langle k|$, with $k,l \in \{0,1,2\}$ and $k < l$. Our protocol is such that in certain cases it is possible to detect the presence of a series of pulses without exciting the detector into the second excited state. This is experimentally realized by trying to detect the presence of a microwave pulse in a transmission line using a transmon qutrit, which serves as a resonantly-activated detector. We require that the detector has not irreversibly absorbed the pulse at the end of the protocol, as witnessed by a non-zero occupation of the second excited state.

Moreover, we employ $N$ Ramsey 
sequences with beam-splitter unitaries $S (\phi)$ to the lowest two energy 
levels. Each of these unitaries are of the form
\begin{equation}
S ( \phi ) = \exp [-i\phi \sigma^{y}_{01}/2]\;. \label{eq:BS}
\end{equation}
or 
\begin{equation}
S ( \phi ) = \mathbb{I}_{01} \cos\frac{\phi}{2} - i \sigma_{01}^{y} \sin\frac{\phi}{2} + |2\rangle\langle 2|, \label{eq:S}
\end{equation}
where $\mathbb{I}_{01} = |0\rangle \langle 0| + |1\rangle \langle 1|$ is the identity operator on the $0-1$ subspace.

The microwave $B$ pulses to be detected are parametrized by a strength $\theta_j$ and a phase $\varphi_j$, and are represented by the unitary

\begin{equation}
B (\theta_{j}, \varphi_{j}) = \exp [-i\theta_{j} \mathbf{n}_{j}\boldsymbol{\sigma}_{12}/2]\;,	  \label{eq:BB}
\end{equation}
where $\mathbf{n}_{j} = (\cos\varphi_j, \sin\varphi_j , 0)$ and $\boldsymbol{\sigma}_{12}=
(\sigma_{12}^{x}, \sigma_{12}^{y}, \sigma_{12}^{z})$, or explicitly
\begin{eqnarray}
	B (\theta_{j}, \varphi_{j}) &=& |0\rangle \langle 0| + \mathbb{I}_{12} \cos\frac{\theta_j}{2} \nonumber \\ 
	&& - i\left(\cos \varphi_{j} \sigma_{12}^{x} + \sin \varphi_{j} \sigma_{12}^{y}\right)\sin\frac{\theta_{j}}{2}, \label{eq:B}
\end{eqnarray}

In matrix form, the $S$ and $B$ operators read
\begin{equation}
S ( \phi ) = 
\begin{pmatrix}	
	\cos\frac{\phi}{2}& - \sin\frac{\phi}{2} & 0 \\
	\sin\frac{\phi}{2} & \cos\frac{\phi}{2} & 0 \\
	0 & 0 & 1
\end{pmatrix},	
\end{equation}
and 
\begin{equation}
	B(\theta_j,\varphi_j ) = 
	\begin{pmatrix}	
		1 & 0 & 0 \\
		0 & \cos\frac{\theta_j}{2} & -ie^{-i\varphi_j} \sin \frac{\theta_j}{2} \\
		0 & -ie^{i\varphi_j} \sin\frac{\theta_j}{2} & \cos\frac{\theta_j}{2}
	\end{pmatrix}	\;. 
	\label{eq:Bmatrix}
\end{equation}

\begin{figure}[h]
	\centering
	\includegraphics[width=1\linewidth]{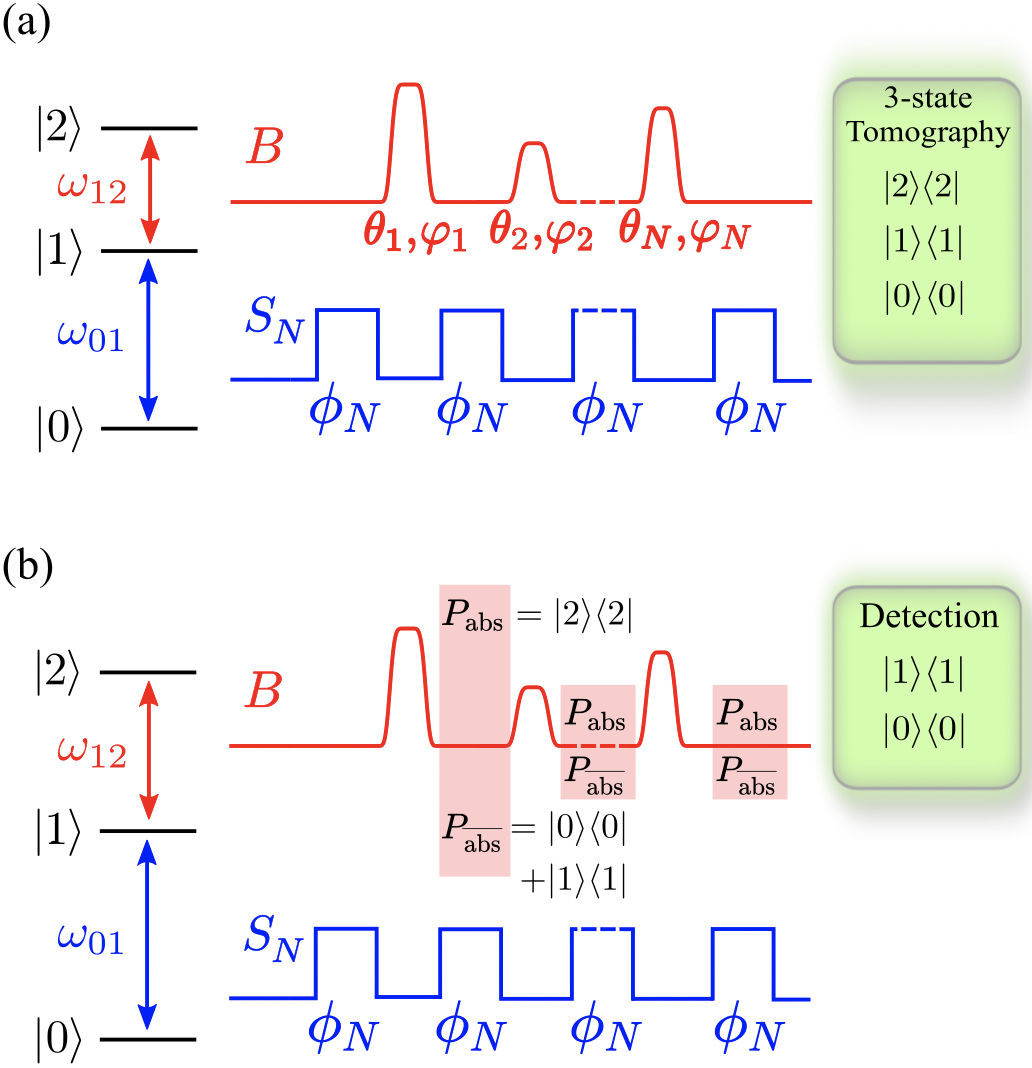}
	\caption{Schematic of (a) the coherent protocol and (b) the projective protocol. Here, $\phi_N = \pi/(N+1)$.  
} 
\label{fig-two_cases}
\end{figure}

The protocol's evolution is governed by a series of $j=\overline{1,N}$ Ramsey sequences, each containing a $B$ pulse with arbitrary $\theta_{j}$ as shown in Fig.~\ref{fig-two_cases}(a). In practice, these unitaries are generated by applying pulses with Hamiltonians $H_{01,j}(t) = -i\hbar[\Omega_{01,j}(t)/2]|0\rangle \langle 1| + \textrm{h.c.}$ and $H_{12,j}(t) = \hbar[\Omega_{12,j}(t)\exp(-i\varphi_{j})/2]|1\rangle \langle 2| + \textrm{h.c.}$ resonant to the $\vert 0 \rangle -\vert 1 \rangle$ and $\vert 1 \rangle - \vert 2 \rangle$ transitions, respectively. The beam-splitter pulses differ only by the times at which they are applied; otherwise, their Rabi frequencies are identical, resulting in $\phi = \int_{-\infty}^{\infty}\Omega_{01,j}(t) dt$. For the $B$ pulses, the Rabi frequencies in general can differ at different $j$'s and therefore we have $\theta_{j} = \int_{-\infty}^{\infty}\Omega_{12,j}(t) dt$.

At the end of the protocol, single-shot measurements (corresponding to projectors $|0\rangle \langle 0|$,  $|1\rangle \langle 1|$, and $|2\rangle \langle 2|$), as well as 
three-level state tomography, can be performed.

We now outline the two main cases below. \bigskip

{\bf Case 1: Absence of $B$ pulses} \bigskip

Here we study the efficiency of the protocol under a general $\phi$. The usual arrangement considered in interaction-free measurements is $\phi \rightarrow \phi_{N}=\pi/(N+1)$ \cite{Kwiat_1995}. From Eq.\,(\ref{eq:BS}), we see that 
$S^{N+1}(\phi_{N}) = \exp (-i \pi \sigma_{01}^{y}/2) =-|0\rangle\langle 1| + |1\rangle\langle 0| + |2\rangle \langle 2|
$. This choice guarantees that for an initial state $|0\rangle$ the resulting state after $N+1$ beam-splitter unitaries is $|1\rangle$, while an initial state $|1\rangle$ would result in a final state $-|0\rangle$. 
When no $B$ pulses are present, the coherent and projective protocols are identical. 

\bigskip

{\bf Case 2: Presence of $B$ pulses} \bigskip

When $B$ pulses are included between each beam-splitter, the evolution of our coherent protocol with $N$ pulses is governed by the string of unitaries
\begin{equation}
\mathbb{U}_{\{N\}} = S(\phi) \prod_{j=1}^{N}\left[B(\theta_{j}, \varphi_{j})S(\phi)\right]\;.\label{eq:U}
\end{equation}
where the product is defined from right to left and the subscript $\{ N\}$ signifies the fact that $\mathbb{U}$ is parametrized by all $\phi$, $\theta_j$, and $\varphi_j$ variables from $j=1$ to $j=N$.
When starting in the state $|0\rangle$, the final state after the application of the full sequence is $\mathbb{U}_{\{N\}} | 0 \rangle$, yielding the probabilities
\begin{align}
	p_{i} = |\langle i| \mathbb{U}_{\{N\}} | 0 \rangle |^2 \label{eq:pureprobabilities}
\end{align}
for $i = \{0,1,2\}$. If the initial state is a density matrix $\rho$, then we have $ \mathbb{U}_{\{N\}} \rho \mathbb{U}^{\dagger}_{\{N\}}$ as the final state and the probabilities are 
\begin{align}
	p_{i} = \mathrm{Tr} \{|i\rangle\langle i| \mathbb{U}_{\{N\}} \rho \mathbb{U}^{\dagger}_{\{N\}} \}\;.
\end{align}

Table~\ref{tab:coh_probs} shows the resulting probabilities and the coherent interaction-free efficiency, $\eta_{\rm c} = p_0/(p_0 + p_2)$, for $N$ = 1, 2, 3, and 4 Ramsey sequences at $\phi = \phi_{N}$, $\theta = \pi$ and $\varphi = \pi/2$, when $\mathbb{U}_{\{N\}}$ acts on the initial state $|0\rangle$.

\begin{table}[H]
	\small 
	\setlength\extrarowheight{2pt} 
	\begin{center}
		\caption{The probabilities $p_{0}$, $p_{1}$, $p_{2}$ and efficiency $\eta_{\rm c}$ for the coherent protocol are shown up to four significant digits for $N = 1, 2, 3$, and $4$. All values are evaluated at $\phi = \phi_{N}$, $\theta = \pi$
			and $\varphi = \pi/2$. The initial state is $|0\rangle$.}
		\begin{tabular*}{\columnwidth}{@{\extracolsep{\fill}} c c c c c}
			\hline\hline
			\textbf{} & \textbf{$N = 1$} & \textbf{$N = 2$} & \textbf{$N = 3$} & \textbf{$N = 4$} \\
			\hline
			\textbf{$p_{0}$} & $0.25$ & $0.8091$ & $0.9983$ & $0.8957$\\
			
			\textbf{$p_{1}$} & $0.25$ & $0.0034$ & $9.721 \times 10^{-4}$ & $0.1041$\\
			
			\textbf{$p_{2}$} & $0.5$ & $0.1875$ & $7.243 \times 10^{-4}$ & $2.505 \times 10^{-4}$\\
			
			\textbf{$\eta_{\rm c}$} & $0.3333$ & $0.8119$ & $0.9993$ & $0.9997$\\
			\hline \hline
		\end{tabular*}
		\label{tab:coh_probs}	
	\end{center}	
\end{table}

We also introduce the following elements of the confusion matrix \cite{our_protocol}: the positive ratio (PR) and the negative ratio (NR). For equal-strength pulses with $\theta_{j}=\theta$, these are defined as $\mbox{PR}(\theta ) = p_{0}(\theta)/(p_{0}(\theta ) + p_{1}(\theta))$ and $\mbox{NR} (\theta )= p_{1}(\theta)/(p_{0}(\theta) + p_{1}(\theta))$. The quantity $\mathrm{FPR}=\mathrm{PR}(\theta = 0)$ is called the false positive ratio and it is a way to characterize the reduction in the confidence of the predictions due to the dark count probability $p_{0}(\theta =0)$ (i.e., positive detection count even in the absence of a pulse).

In contrast, for the standard projective case \cite{Elitzur_1993},
as usually implemented in optics, the POVM measurement operators after each application of the $B$ pulse are 

\begin{eqnarray}
P_{\overline{\rm abs}} &=& \mathbb{I}_{01}=|0\rangle \langle 0| + |1\rangle \langle 1| \label{eq:Pnonabs}\;,\\
P_{\rm abs} &=& |2\rangle \langle 2|,
\end{eqnarray}
where $P_{\rm{abs}}$ is the projector corresponding to an absorption event, while $P_{\overline{\rm abs}}$ describes the situation where absorption did not occur. Fig.~\ref{fig-two_cases}(b) illustrates this detection scheme for a protocol with $N$ steps. 
Note that by including the $B$ pulse we can define the POVM measurement operators $M_{\rm abs}=P_{\rm abs}B$ and $M_{\overline{\rm abs}}=P_{\overline{\rm abs}}B$, satisfying the completeness property $M_{\rm abs}^{\dag}M_{\rm abs} + M_{\overline {\rm abs}}^{\dag}M_{\overline{\rm abs}} = \mathbb{I}_{3}$, where $\mathbb{I}_{3}$ is the $3\times 3$ identity matrix.

In this protocol, it is useful to define two probabilities: the probability of detection $p_{\rm{det}}$ and the probability of absorption $p_{\rm{abs}}$ \cite{Kwiat_1995}. For $N$ $B$ pulses, these probabilities can be obtained by applying Wigner's generalization of Born's rule \cite{Wigner_1963}

\begin{align}
	p_{\rm det} = \left| \langle 0|  \mathbb{X}_{\{N\}} |0\rangle \right|^2\;,  \label{eq:det}
\end{align}

\begin{align}
	p_{\rm abs} = \sum_{j = 1}^{N} \left| \langle 2|  B(\theta_{j}, \varphi_{j})\mathbb{X}_{\{j-1\}}
|0 \rangle   \right|^{2} \;, \label{eq:abs}
\end{align}
where the string of operators
\begin{equation}
\mathbb{X}_{\{j\}}=S(\phi )\prod_{i=1}^{j} \left[ P_{\overline{\rm abs}}B(\theta_{i}, \varphi_{i}) S (\phi )\right],
\end{equation}
with the convention $\mathbb{X}_{0}=\mathbb{I}_{3}$ (the $3 \times 3$ identity matrix), now plays the role of the $\mathbb{U}_{N}$ unitary from the coherent case, see Eq.\,(\ref{eq:U}).

 One can readily verify that $p_{\rm det}$ is a product of probabilities: the probability of detection on the state $|0\rangle$ when applying $S$ the $(N+1)$th time, multiplied by the probability that the wavefuction did not collapse to $|2\rangle$ in any of the previous $N$ detection steps. Similarly, $p_{\rm abs}$ is a sum of probabilities, each obtained as a product  between the state-$|2\rangle$ probability after applying $B$ in step $j$ and the probability that the wavefunction did not collapse to $|2\rangle$ in any of the previous $j-1$ detection steps.  

For mixed states, these expressions generalize immediately as 

\begin{align}
	p_{\rm det} = \mathrm{Tr} \{|0\rangle\langle 0| \mathbb{X}_{\{N\}}  \rho \mathbb{X}^{\dag}_{\{N\}}
	\} \;,
\end{align}	
\begin{align}
	p_{\rm abs} = \sum_{j = 1}^{N}  \mathrm{Tr} \{|2\rangle\langle 2|  B(\theta_{j}, \varphi_{j}) 
\mathbb{X}_{\{j-1\}}\rho \mathbb{X}^{\dag}_{\{j-1\}}  B^{\dag}(\theta_{j}, \varphi_{j})  \}\;,
\end{align}
where $\rho$ is the initial state. 
As we will see later, in real systems under decoherence, the operators $S$ and $B$ will also be modified accordingly.

Table~\ref{tab:proj_probs} shows the resulting probabilities and the projective interaction-free efficiency, defined as $\eta = p_{\rm det}/(p_{\rm det} + p_{\rm abs})$, for  $N$ = 1, 2, 3, and 4, at $\phi = \pi/(N+1)$, $\theta = \pi$ and $\varphi = \pi/2$, starting with the initial state $|0\rangle$.
Comparing the efficiencies from Table I and II, we note a clear advantage of the coherent interaction-free detection protocol over the projective one, with the coherent efficiency $\eta_c$ already exceeding 0.999 for $N=3$.

\begin{table}[H]
	\small 
	\setlength\extrarowheight{2pt} 
	\begin{center}
		\caption{The probabilities $p_{\rm det}$ and $p_{\rm abs}$, along with the efficiency $\eta$ for the projective protocol, are shown up to four significant digits for $N = 1, 2, 3,$ and $4$. All values are evaluated at $\phi = \phi_{N}$, $\theta = \pi$, and $\varphi = \pi/2$. The initial state is $|0\rangle$.}
		\begin{tabular*}{\columnwidth}{@{\extracolsep{\fill}} c c c c c}
			\hline\hline
			\textbf{} & \textbf{$N = 1$} & \textbf{$N = 2$} & \textbf{$N = 3$} & \textbf{$N = 4$} \\
			\hline
			\textbf{$p_{\rm det}$} & $0.25$ & $0.4219$ & $0.5308$ & $0.6054$\\
			
			\textbf{$p_{\rm abs}$} & $0.5$ & $0.4375$ & $0.3781$ & $0.3307$\\
			\
			\textbf{$\eta$} & $0.3333$ & $0.4909$ & $0.5840$ & $0.6468$\\
			\hline\hline
		\end{tabular*}
		\label{tab:proj_probs}	
	\end{center}	
\end{table}

Similarly to the coherent case, we can also introduce the positive and negative ratios of projective interaction-free measurements by replacing $p_0$ and $p_{1}$ with $p_{\rm det}$ and $1-p_{\rm abs}-p_{\rm det}$, respectively.

A few observations are in place at this point. One is that the three-state model with projection operators is able to emulate exactly the physics of an ultrasensitive object placed in one arm of a chain of Mach-Zehnder interferometers, as usually studied in quantum optics. The only difference is that in the latter case the measurement is destructive, while in our case the projector $|2\rangle \langle 2|$ is a von Neumann non-demolition operator. However, this is not a serious issue: one can connect the $|2\rangle \langle 2|$ detector to an instrument that simply switches off the experiment. Another way of realizing this in circuit QED is by using a phase qubit with the states $|0\rangle$ and $|1\rangle$ localized in one of the wells of the washboard potential and with the state $|2\rangle$ close enough to the barrier that switching into the running state occurs by tunneling with some probability \cite{Paraoanu_2006}.

Another observation is that in the projective case, the probability $p_{\rm abs}$ is calculated immediately after the last $B$ pulse, while in the coherent case, all probabilities are calculated after the last beam-splitter $S$. However, the last $S$ acts only on the $0-1$ subspace; therefore, the probability of state $|2\rangle$ remains invariant under its action. We can thus perform a point-to-point fair comparison of the two protocols.

\section{Results in the large-$N$ limit} \label{sec_3}

In this section, we derive approximate expressions for the probability amplitudes when our protocol is subjected to a large number of Ramsey sequences $N$. We also explore the lower limit to the $B$-pulse strength which can still give rise to high enough interaction-free detection efficiency.

\subsection{Analytical results} \label{sec_3a}

The coherent interaction-free protocol has been reported to yield high efficiencies when the number $N$ of consecutive Ramsey sequences is large \cite{our_protocol}. Here, we present a detailed analysis of this case using analytical tools. 
Let us consider the beam-splitter unitary $S(\phi_N) =\exp(-i\phi_N \sigma_{01}^y/2)$ from Eq.\,(\ref{eq:BS}), where  $\phi_N=\pi/(N+1)$ is the beam-splitter strength that presents constructive interference on state $|1\rangle$ in the absence of $B$ pulses.  For the $B$-pulse unitary we choose $B(\theta) = B(\theta, \pi/2) =\exp(-i\theta \sigma_{12}^y/2)$; in other words, 
we take all $\varphi_j=\pi/2$ for simplicity [see Eq.\,(\ref{eq:BB})].

We start with the observation that $\mathbb{U}_{\{N\}} = S (\phi_N)\left[B(\theta)S (\phi_N) \right]^N = \left[S (\phi_N)B(\theta ) \right]^{(N+1)}B^{-1}(\theta )$. Since $B(\theta)$ does not act on the ground state, it follows that 
$B^{-1}(\theta )|0\rangle = |0\rangle$, and therefore the final state can be obtained as
\begin{equation}
	\mathbb{U}_{\{N\}}|0\rangle = \left[ S (\phi_N) B(\theta) \right]^{N+1}  |0\rangle\;.
\end{equation}

Next, our goal is to obtain an approximate spectral decomposition of the matrix $S (\phi_N) B(\theta)$ in the limit $\phi_{N} \ll 1$ and $\cos (\theta /2) \ll 1$. The details of this calculation are presented in Appendix~\ref{Appendix_A}. We find the eigenvalues $1, e^{-i\theta/2}, e^{i \theta /2}$ with corresponding eigenvectors appearing as columns in the diagonalizing matrix $M$

\begin{displaymath}
	M = \left( \begin{array}{ccc}
		1 & \frac{1}{2}\phi_N & \frac{1}{2}\phi_N \\ 
		\frac{1}{4}\phi_N & 2i  \sin \frac{\theta}{4}e^{i \theta /4} & -2i \sin \frac{\theta}{4}e^{-i \theta /4}  \\ 
		\frac{1}{4}\phi_N \cot\frac{\theta}{4} &-2  \sin \frac{\theta }{4} e^{i \theta /4} &  -2  \sin \frac{\theta}{4}e^{-i \theta /4}
	\end{array} \right) .
\end{displaymath}
We can then obtain the matrix  $\left[ S (\phi_N) B(\theta) \right]^{N+1}$ as
\begin{displaymath}
	\left[ S (\phi_N) B(\theta) \right]^{N+1} = M \cdot \left( \begin{array}{ccc}
		1 & 0 & 0 \\ 
		0 & e^{-i (N+1)\theta/2} & 0 \\ 
		0 & 0 & e^{i (N+1)\theta/2}
	\end{array} \right) \cdot M^{-1}.
\end{displaymath}

		Consider now the final state written in the form
		$c_{0}|0\rangle + c_{1}|1\rangle + c_{2}|2\rangle  = \mathbb{U}_{\{N\}}|0\rangle$. Using the results above, after some algebra, we obtain
		\begin{eqnarray}
			c_{0} &=& 1 - \frac{1}{2} \left(\frac{\phi_{N}}{2}\right)^2 \frac{1}{\sin^{2}\frac{\theta}{4}} \sin^2 \frac{(N+1)\theta}{4}\;, \label{eq:c0approx}\\
			c_{1} &=& \frac{\phi_N}{2} \frac{1}{\sin \frac{\theta}{4}}\cos \frac{N\theta}{4} \sin \frac{(N+1)\theta}{4}\;, \label{eq:c1approx}\\
			c_{2} &=&\frac{\phi_N}{2} \frac{1}{\sin \frac{\theta}{4}}\sin \frac{N\theta}{4} \sin \frac{(N+1)\theta}{4}. \label{eq:c2approx}
		\end{eqnarray}
		One can see just by inspection that the wavefunction is correctly normalized up to fourth order in $\phi_N$.

		The  detection efficiency of the coherent protocol is given by
		\begin{equation}
			\eta_{\rm c}=\frac{p_0}{p_0 + p_2} = \frac{|c_0|^2}{|c_0|^2 + |c_2|^2},
		\end{equation}
		which, using the results above, can be evaluated to
		\begin{equation}
			\eta_{\rm c} \approx 1- \frac{\phi_{N}^2}{16\sin^2\frac{\theta}{4}} \left[ \cos\frac{\theta}{2}-\cos \left( \frac{N\pi}{2} + \frac{\pi}{4}\right) \right]^2\;. \label{eq:scale}
		\end{equation}
		This shows that the efficiency approaches 1 in an oscillatory way, exactly as observed in the numerical simulations.

		These results allow us to obtain even deeper insights into the mathematics of our protocol. In the asymptotic large-$N$ limit, and if $\theta \gg 1/N$, we can completely neglect even the first order terms in $\phi_N/2$. The coefficients of the final wavefunction become $c_0 \approx 1$, $c_{1} \approx c_{2}  \approx 0$, so the protocol achieves nearly perfect efficiency. To understand why this is the case, we can calculate $\mathbb{U}_{\{N\}} = \left[S (\phi_N)B(\theta ) \right]^{(N+1)}B^{-1}(\theta )$ in this limit, obtaining
		\begin{equation}
			\mathbb{U}_{\{N\}} \approx \left( \begin{array}{cc} 1 & O_{1\times2} \\ O_{2\times1} & \mathcal{B}(N\theta) \end{array}
			\right), \label{eq:matrx}
		\end{equation}
		where $O_{n1 \times n2}$ is a null matrix of dimension $n_1 \times n_2$ and $\mathcal{B} (\theta)$ is the submatrix in the $1-2$ subspace from Eq.\,(\ref{eq:Bmatrix}) with $\theta_j = \theta$ and $\phi_j = \pi /2$.
		 Asymptotically, the evolution is approximated as a rotation with an angle $N\theta$ in the $1-2$ subspace. When we apply this operator to an initial state $|0\rangle$, the state of the system remains unaltered with very high probability. This is straightforward quantitative evidence that in the coherent interaction-free protocol, the state of the system at large $N$ mostly does not evolve, and still one can detect the presence of a $B$ pulse with very high probability. 

\subsection{Discussion: limits on $\theta$}  \label{sec_3b}

\begin{figure}[h]
	\centering
\includegraphics[width=0.8\linewidth]{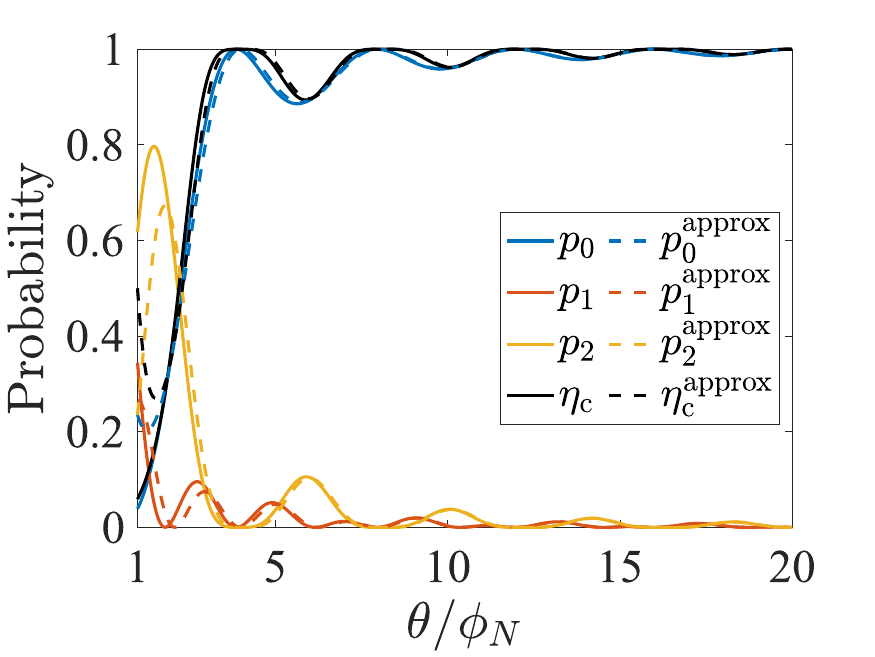}
	\caption{The probabilities and efficiency of our protocol vs $\theta/\phi_N$ for $N = 25$. The exact numerical values are represented by solid lines, whereas the approximations are represented by dashed lines. We see that for $\theta/\phi_N \ge 4$, the ground state probability $p_0$ and efficiency $\eta_{\rm c}$ approach unity.}
\label{fig-largeN}
\end{figure}

Next, we obtain the minimum $B$-pulse strength for which the above approximate treatment works appropriately. In general, for small $B$-pulse strength (e.g., $\theta \approx \phi_N$) the coherent interaction-free detection protocol may not result in an efficient detection. We address this issue numerically by examining the probabilities of the final state as a function of the ratio $\theta/\phi_N$, as shown in Fig.~\ref{fig-largeN}. In this figure, the exact numerical results based on evolving the system according to Eq.\,(\ref{eq:pureprobabilities}) are compared with the approximate results from Appendix~\ref{Appendix_A}  based on the treatment above. Very similar results are obtained when using the simpler expressions from the previous subsection.  
We have checked numerically that the variation of probabilities $p_0$, $p_1$, $p_2$ versus $\theta/\phi_{N}$ is not very sensitive to the value of $N$; that is, the probability profiles remain almost the same as in Fig.~\ref{fig-largeN} for any arbitrary value of $N$. 

We notice that $p_0$ reaches close to $1$ at $\theta/\phi_N \simeq 4$, and thereafter remains close to $1$ for $4 \leq \theta/\phi_{N} \leq 4N$. At $\theta/\phi_{N}=4(N+1)$, which means $\theta =4\pi$, it drops again to zero, reflecting the $4\pi$ periodicity of the system (see also \cite{our_protocol} for the experimental observation of this effect). Thus, $\theta \simeq 4\phi_N$ is the minimum value of $B$-pulse strength that gives a highly efficient interaction-free detection (see the solid blue curve in Fig.~\ref{fig-largeN}). 

We can understand where this value comes from by examining the approximate solutions Eqs.\,(\ref{eq:c0approx})--(\ref{eq:c2approx}): we see that at $\theta = 4\phi_N$ the last sine function in these expressions becomes zero. 
Further, we see that the lower and upper limits of $\theta/\phi_{N}$, namely $4$ and $4N$, respectively, mark the boundaries of the plateau later discussed in Fig.~\ref{fig-threshold} and 
observed experimentally in Fig. 6 in  Ref.~\cite{our_protocol}, extending from $\theta = 4\phi_{N}$
to $\theta=4 N \phi_{N} = 4\pi-4\phi_{N}$. The width of the plateau is $2(2\pi-2 \phi_{N})$, which we have verified also by direct comparison with the numerical data from Fig.~\ref{fig-threshold}. This width is therefore zero for $N=1$ ($p_0$ attains its maximum value close to 1 only at $\theta =2\pi$ and has a downward trend as $\theta$ exceeds $2\pi$). The limits (or the width) of these plateaus of highly efficient interaction-free detection are further attributed to the periodicity of the protocol in $\theta$ with a period $4\pi$. 

In the next sections, we will be more interested in exploring the lower limit on the $B$-pulse strength, which can give rise to near-unity interaction-free detection efficiency. It is also noteworthy that the solid curves in Fig.~\ref{fig-largeN} result from numerical simulations without considering the large-$N$ approximation. Thus, the bounds on $\theta/\phi_N$ obtained here represent a general characteristic of our protocol.

\section{Information in coherent interaction-free detection} \label{sec_4}

The effect of $B$ pulses on the success probability and efficiency of each protocol is more thoroughly explored in this section, with the goal of providing further insights into how information on the presence of the pulses is acquired during the protocol. We begin by considering $B$ pulses with equal strengths and study the behavior of the success probabilities of both protocols at different $B$-pulse strengths $\theta$ and $N$ Ramsey sequences. Further, we explore the successive probabilities at different $N$ of the system evolutions for both protocols with $B$ pulses of strength $\theta = \pi$. Finally, we provide an analysis based on Fisher information, which demonstrates that the precision at which we can determine a small $\theta$ obeys the Heisenberg scaling.

\subsection{$B$ pulses with equal strengths}\label{sec_4a}

\begin{figure}[h]
	\centering
	\includegraphics[width=0.9\linewidth]{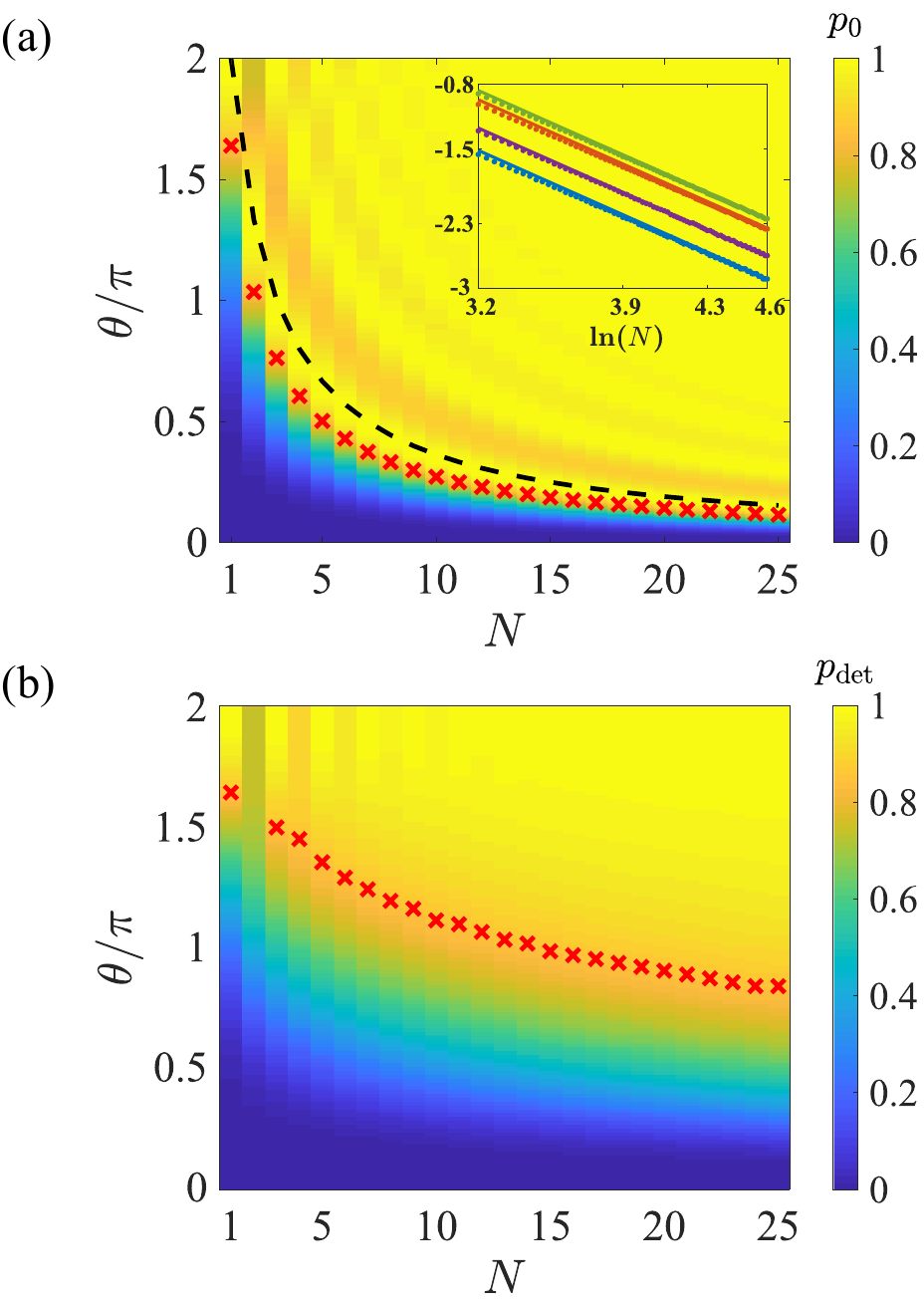}
	\caption{The surface map for (a) $p_0$ and (b) $p_{\rm det}$ as functions of $\theta$ and $N$ in an ideal case 
		of identical $B$ pulses. Red `x' markers correspond to the threshold values 
		of $\theta$ at $p_{0} \ge 0.85$ and the dashed black line in (a) denotes the values $\theta = 4\phi_{N}$. The inset in (a) represents $\ln(\theta/\pi)$ with circle symbols  taken at four threshold values $p_{0} \ge 0.25$ (blue), $p_{0} \ge 0.5$ (magenta), $p_{0} \ge 0.85$ (red), and $p_{0} \ge 0.95$ (green).  For this analysis, the range of $N$ is extended to $N \in [25,100]$. The solid lines of the same color are the lines of best fit after taking the natural logarithm of the power law $aN^{-1}$, i.e., $\ln(a) - \ln(N)$.}
	\label{fig-threshold}
\end{figure}

While the coherent protocol generally has a higher success probability than the projective protocol, it is useful to see just how they differ at various $N$ for different fixed $B$-pulse strengths $\theta$. 
Here, we consider the success probability profiles of each protocol for various values 
of $N \in [1,25]$ as a function of $\theta$, shown in 
Fig.~\ref{fig-threshold}, with optimal beam-splitter strengths $\phi = \phi_{N}$. For a given $N$, all $B$ pulses are of the same strength $\theta$, varying linearly between $[0,2\pi]$. Both $p_0$ and $p_{\rm det}$ are symmetric about $\theta=2\pi$ and, as expected, gradually rise from $0$ to a maximum value with increasing $\theta$, tending to stay high and forming a plateau with a noticeably wavy structure for $p_{0}$. This plateau gets wider with increasing $N$. 

The same $p_0$ can also be recognized as a quantitative measure of the success probability of the interaction-free measurement. Thus, the widening of the $p_0$ plateau (close to $1$) for higher values of $N$ demonstrates that systems with multiple $B$ pulses can perform interaction-free detection with very high efficiency. Beyond a threshold $\theta$, this efficiency becomes almost independent of $\theta$.
This threshold is represented by red markers 
in Fig.~\ref{fig-threshold}(a), plotted on top of the $p_0$
distribution as a function of $N$ and 
$\theta$, corresponding to an ideal case of identical 
$B$ pulses. Data shown with red markers in fact correspond to 
a minimum $\theta$ satisfying $p_0\geq 0.85$. 

The dashed black line represents $p_0$ at $\theta = 4\phi_{N}$, and the inset is the log-log plot of the populations at thresholds $p_{0} \ge 0.25$ (blue), $p_{0} \ge 0.5$ (magenta), $p_{0} \ge 0.85$ (red), and $p_{0} \ge 0.95$ (green) when considering  $N \in [25,100]$. The circles indicate $\ln(\theta/\pi)$ at these thresholds, and the solid lines are the best fits of the form $\ln{(aN^{-1})}$. We find that the coefficients are approximately $a =$ 1.7, 2.2, 2.9, and 3.3 at thresholds $p_{0} \ge 0.25$, $0.5$, $0.85$, and $0.95$, respectively.

In Fig.~\ref{fig-threshold}(b), we note that the threshold $p_{\rm det} \ge 0.85$ is considerably higher in $\theta$ than the corresponding threshold of the coherent protocol. In fact, for $N = 2$, the threshold is not even reached, as can be seen from the lack of a marker. 
Thus, the coherent protocol generally has higher success probabilities over a wider range of $\theta$, even at small $N$.

The $N^{-1}$ scaling seen in Fig.~\ref{fig-threshold}(a) can also be obtained from Eq.\,(\ref{eq:c0approx}) in the following manner. A fixed value of $p_{0}$, not too close to 1 (corresponding to the chosen threshold), is obtained at relatively low values of $\theta$ by fixing the ratio $\theta /\phi_{N}$ at a constant value. This immediately results in the scaling $\theta \sim N^{-1}$ observed numerically. If the measurement of $p_0$ is utilized as a way of measuring $\theta$, this yields Heisenberg-scaling precision. We will confirm this result later in this section when analyzing the Fisher information. 

\subsection{Successive probabilities of detection and absorption} \label{sec_4b}
\begin{figure}[ht]
	\centering
	\includegraphics[width=1 \linewidth,keepaspectratio=true]{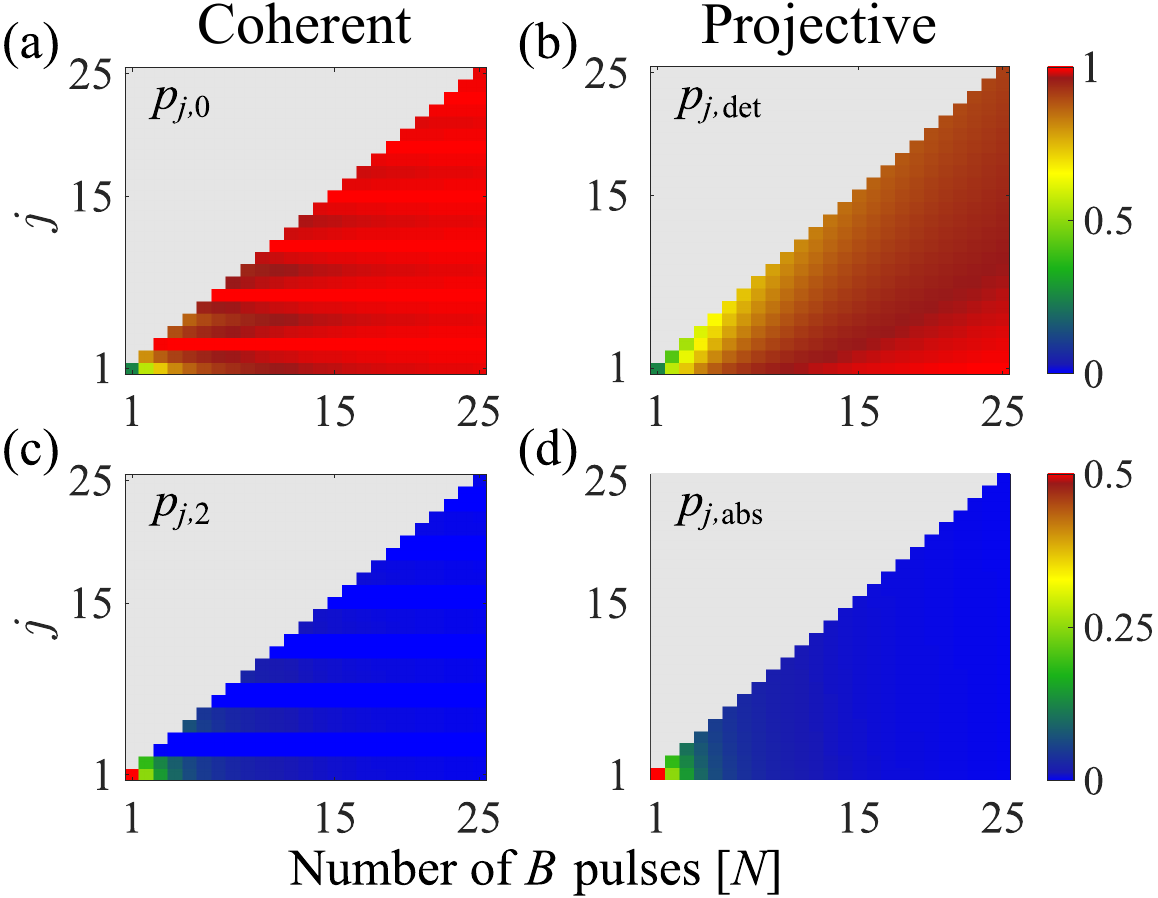}
	\caption{Probabilities of success for successive $B$-pulse implementations: (a) $p_{j,0}$ for the coherent protocol and (b) $p_{j, {\rm det}}$ for the projective protocol, with values in the range $[0,1]$ indicated by the colorbar. Plots (c) and (d) show the probabilities of absorption in the coherent and projective protocols, respectively. All $B$ pulses are of strength $\theta = \pi$.}
	\label{fig:fig6}
\end{figure}
Here we further develop insights into the coherent interaction-free and projective measurement-based protocols by looking at the detailed map of successive probabilities of detection and absorption at the end of each Ramsey sequence $j$ when subjected to $B$ pulses of strength $\theta = \pi$. These probabilities are denoted respectively by $p_{j,0}$ and $p_{j,2}$ for the coherent case, and $p_{j,{\rm det}}$ and $p_{j,{\rm abs}}$ for the projective case. At the end of the sequence ($j = N$), using the previous notation, we have
$p_{N,0} \equiv p_0$, $p_{N,2} \equiv p_2$, etc. Thus, these maps show how the probability of occupation of the three levels evolves with successive Ramsey sequence implementations up to $N$. 	
 
Figures~\ref{fig:fig6}(a) and~\ref{fig:fig6}(b) present the ground state probability plotted for $j \in [1,N]$, $N \in [1,25]$. 
 As $N$ increases, $p_{j,0}$ tends to $1$ very rapidly, as shown by the bright red in the color map, while $p_{\rm det}$ manages to exceed $0.9$ marginally for $N=23$. Implementing the first Ramsey sequence ($j=1$) results in the same values $p_{1,0}=p_{1,\mathrm{det}}$
 for any arbitrary $N$. This is due to the fact that the $j=1$ coherent and projective sequences do not differ in any fundamental way when performing a POVM analysis \cite{our_protocol}.

In the coherent protocol, $p_{j,0}$ increases for $j\geq2$ and then oscillates with $j$ in the range $[0.85, 0.999]$, which further subsides for large $N$. Typically, for large $N$, say $N=25$ in the coherent protocol, the system tends to stay in the initial state $|0\rangle$ with a very high probability ($> 0.99$) throughout the sequence.
Higher values of $p_{j,\rm det}$ at large $N$ with small values of $j$ correspond to higher ground state occupancy for the first few steps, which should not be mistaken as higher probability of interaction-free detection.

Similarly, Figs.~\ref{fig:fig6}(c) and~\ref{fig:fig6}(d) present the probability of the second excited state $p_{j,2}$ and the probability of $B$-pulse absorption $p_{j,\rm abs}$ at the end of the $j$th Ramsey sequence
implementation for a given $N$ in the coherent interaction-free and projective measurement protocols, respectively. Mirroring the features of the $p_{j,0}$ map, the map of $p_{j,2}$ also exhibits a pattern of oscillations with $j$, where the bright blue color corresponds to $p_{j,2}$ values as low as $0.01$ and the dark blue color corresponds to slightly larger values.

\subsection{Fisher information of the protocols} \label{sec_4c}

Since our protocol is remarkably efficient, it can, in principle, be used to provide an estimate for the $B$-pulse strength $\theta$. Here, we study the associated quantum Fisher information of our protocol and that of the projective case at $\theta = 0$ to determine which quantum limits are reached. We imagine two situations: one in which all three probabilities are used for evaluating $\theta$, and another in which only two probabilities, which make up the efficiency, are used.

The Cram\'er-Rao bound states that

\begin{equation}
	\text{Var}\left(\hat{\theta} \right) \geq \frac{1}{\text{QFI} (\theta)}, 
	\label{eq:CR-bound}
\end{equation}
where the variance of the parameter $\theta$ is bounded by the Fisher information of the parameter QFI($\theta$) \cite{Wander}. 
Moreover, the Fisher information is defined as 

\begin{equation}
	\text{QFI} (\theta) = \sum_{i=0,1,2}\frac{[\partial_{\theta} (p(i|\theta))]^{2}}{p(i|\theta)}\;.
	\label{eq:Fisher}
\end{equation}

	Thus, the Fisher information of the coherent protocol is 
	\begin{equation}
		{\rm QFI}_{\rm c} = \frac{(\partial p_0/\partial \theta)^{2}}{p_0} + \frac{(\partial p_1/\partial \theta)^{2}}{p_1} + \frac{(\partial p_2/\partial \theta)^{2}}{p_2}\;,
	\end{equation}
	and for the projective protocol, it is 
	\begin{align}
		{\rm QFI}_{\rm proj} = \frac{(\partial p_{\rm det}/\partial \theta)^{2}}{p_{\rm det}} + \frac{(\partial p_{\rm abs}/\partial \theta)^{2}}{p_{\rm abs}} + \nonumber \\ \frac{[\partial (1 - p_{\rm det} - p_{\rm abs})/\partial \theta]^{2}}{(1-p_{\rm det} - p_{\rm abs})}\;. 
	\end{align}
	The Fisher information of the efficiency of each protocol characterizes how sensitive the efficiency is with respect to a variable, e.g.,  $\theta$. 
	Explicitly, this is
	
	\begin{equation}
		{\rm QFI}_{\eta_{\rm (c)}} = \frac{(\partial \eta_{\rm (c)}/\partial \theta)^{2}}{\eta_{\rm (c)}} + \frac{[\partial (1-\eta_{\rm (c)})/\partial \theta]^{2}}{1- \eta_{\rm (c)}}\;,
	\end{equation}
	or, more compactly, 
	\begin{equation}
		{\rm QFI}_{\eta_{\rm (c)}} = 
		\frac{1}{\eta_{\rm (c)}(1-\eta_{\rm (c)})}
		\left(\frac{\partial \eta_{\rm (c)}}{\partial \theta}\right)^2\;.
	\end{equation}

\begin{figure}[ht]
			\centering
			\includegraphics[width=1\linewidth,keepaspectratio=true]{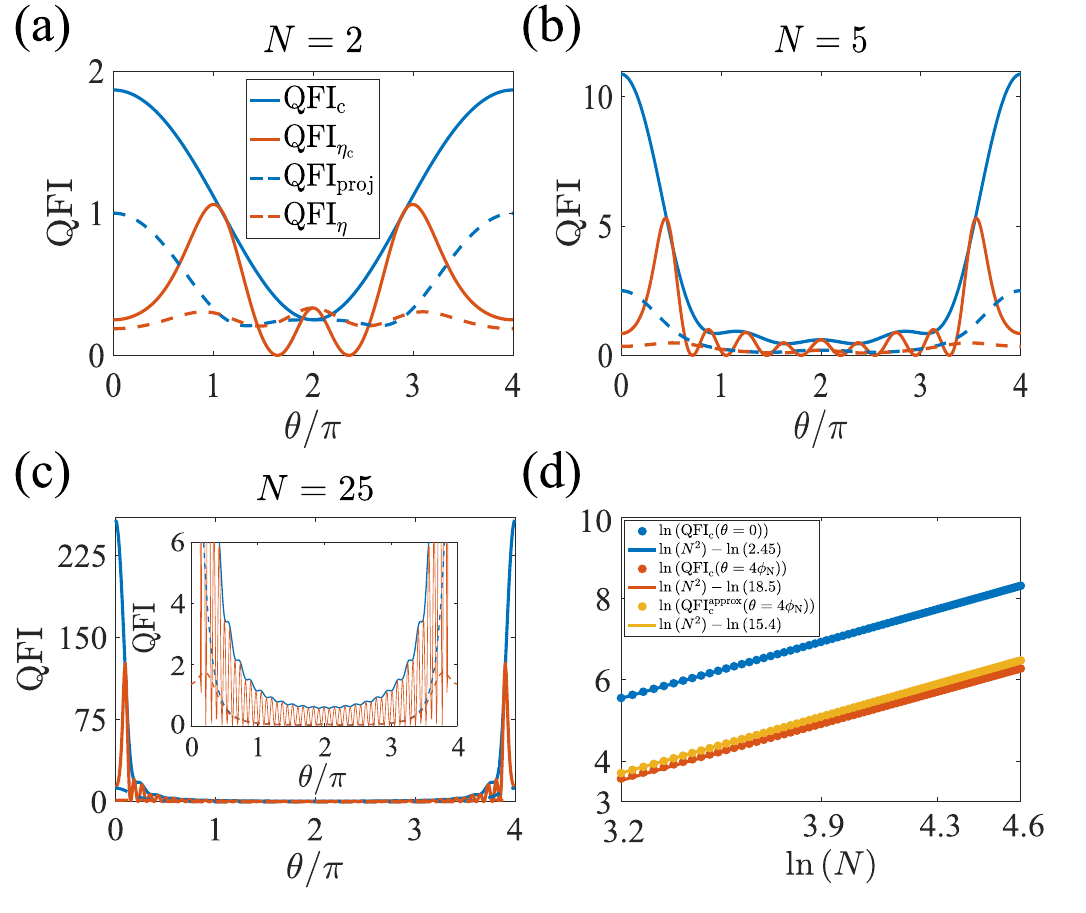}
			\caption{
				The quantum Fisher informations for  $N$ = 2 (a), 5 (b), and 25 (c) for both coherent and projective cases. Panel (c) also contains an inset showing a different resolution of the QFI. Each case is evaluated using the Fisher informations $\rm QFI_{\rm c}$, $\rm QFI_{\eta_{\rm c}}$, $\rm QFI_{\rm proj}$, and $\rm QFI_{\eta}$. (d) Log-log plot of $\rm{QFI_c}$ at $\theta = 0$ (blue circles) along with the exact solution of $\rm{QFI_c}$ at $\theta = 4\phi_N$ (red circles) and the corresponding approximate solution (yellow circles) using the probability amplitudes derived in Sec.~\ref{sec_3}. Each case is evaluated in the range $N \in [25, 100]$. The blue, red, and yellow solid lines are the corresponding best-fit lines.}
			\label{fig:qfi}
		\end{figure}

		As can be seen in Fig.~\ref{fig:qfi}, the Fisher informations $\rm{QFI}_{\rm c}$ and $\rm{QFI}_{\rm proj}$ each have maximum values at $\theta = 0$ and $\theta =4\pi$, regardless of $N$. In general, $\rm{QFI}_{\eta}$ and $\rm{QFI}_{\eta_{\rm c}}$ do not reach their maxima exactly at $\theta = 0$ or $4\pi$, and these maxima converge to $\theta = 0$ as $N \to \infty$. This shows that the most interesting situation for determining $\theta$ with high precision occurs at small values (or values near $4\pi$). We can also see that this should indeed be the case by examining Fig.~\ref{fig-threshold}(a): there, for $N \gg 1$ the maximum variation of $p_0$ ---  which is metrologically useful ---  occurs at very low values of $\theta$.
		
		Taking the limit of ${\rm QFI}_{\textrm{c}}$ and $\rm{QFI}_{\rm proj}$ as $ \theta \rightarrow 0$, we observe that the projective case is precisely $N/2$ and that for the coherent case the power law fitting is $0.42N^{2}$. 
 Hence, the projective protocol reaches the standard quantum limit (SQL) and the coherent case approaches the Heisenberg limit.  Similarly, $\rm QFI_{\eta} =$  $0.1N + 0.05$ and $\rm QFI_{\eta_{\rm c}} =$ 
		$0.024N^{2}$ (for points larger than $N = 35$) as $ \theta \rightarrow 0$, so the SQL and Heisenberg limit are also respectively reached for the Fisher information of the efficiencies. 
	
	For completion, we also investigate the quantum Fisher information at $\theta = \pi$, where the protocol tends to be less sensitive compared to small $\theta$, due to the formation of plateaus of probabilities (see discussion in Sec.~\ref{sec_3b} and Sec.~\ref{sec_4a}). At $\theta = \pi$ and for $N \in [200, 10^3]$, $\rm{QFI_{proj}}$ monotonically decreases following approximately the power law $2.5N^{-1}$; similarly, $\rm{QFI}_{\eta}$ monotonically decreases, except at $N = 1$,
also following approximately the power law $2.5N^{-1}$. $\rm{QFI_c}$ oscillates in an underdamped fashion with respect to $N$ and converges to approximately $1.2$ as $N \to \infty$. $\rm{QFI}_{\eta_{\rm c}}$ at $\theta = \pi$ also oscillates in an underdamped fashion and converges to approximately $0.62$ as $N \to \infty$. 

			Next, we use the approximate expressions for the final state coefficients ($c_0$, $c_1$, $c_2$) obtained in the limit of large $N$ [see Eqs.\,(\ref{eq:c0approx})--(\ref{eq:c2approx}) or Eqs.\,(\ref{Eq:c0})--(\ref{Eq:c2})] and obtain the expressions for the quantum Fisher information as a function of $\theta$ and $N$. To this end, we again use the ratio $\theta /\phi_{N}$.
			
			Based on the results in Sec.~\ref{sec_3b}, we know that we can approach small values of $\theta$ down to $\theta /\phi_{N} \simeq 4$, and the approximate equations from Sec.~\ref{sec_3a} will still be valid. Thus, we can analytically calculate the Fisher informations under the approximations $\theta /\phi_{N} \ll N$ and $\phi_N \ll 1$. For 
			$\mathrm{QFI}_{\eta_{\rm c}}$, we obtain
			\begin{equation}
			\mathrm{QFI}_{\eta_{\rm c}} \approx  \frac{4 N^2}{\pi (\theta /\phi_{N})^2}\frac{\left[1-\cos(N\pi /2 + \pi/4 )\right]^2}{(\theta /\phi_{N})^2 - \left[1-\cos(N\pi /2 + \pi/4 )\right]^2}.
			\end{equation}
			
	This shows that $\mathrm{QFI}_{\eta_{\rm c}}$ scales as $N^2$ for small values of $\theta$ (on the order of $\phi_N$). One can also see that this scaling does not hold if $\theta /\phi_{N}$ becomes on the order of $N$ (or, in other words, $\theta$ becomes comparable to $\pi$), as also seen numerically.
			In the case of $\textrm{QFI}_\textrm{c}$, we can perform a similar analysis, with the result  ${\rm QFI}_{\rm c} \propto N^2$ at large $N$. The final expressions are too cumbersome to be reproduced here; instead, we will make some further observation based on numerical results. 
			
			With increasing $N$, the parameter $\theta$ to be estimated decreases with $N$, while the variance in its estimation decreases with $N^2$. Further, it is seen that for an arbitrarily chosen fixed value of $\theta$, the ${\rm QFI}_{\rm c}$ saturates to a constant value for large $N$, which is inversely proportional to the value of $\theta$. Fig.~\ref{fig:qfi}(d) shows the $N^{2}$ proportionality of both $\rm QFI_{\rm c}(\theta = 0)$ and $\rm QFI_{\rm c}$ $(\theta = 4\phi_N)$ along the corresponding best-fit lines. The latter was explored due to $4$ being the lowest value  of $\theta/\phi_{N}$ where the efficiency is high, as seen in Fig.~\ref{fig-largeN}. Thus, we see if the Heisenberg limit is reached for these choices of $\theta$.

To get some intuition of why the coherent protocol performs better than the projective one, we examine the recursion relations in Appendix~\ref{Appendix_B}. We can see that in the projective protocol the information about $\theta$ contained in the amplitude of state $|2\rangle$ is erased at every application of $P_{\overline{\rm abs}}$, and what is being measured at every step and retained in the ground- and first-excited state amplitudes is $\approx \cos (\theta /2)$. In particular, for $\theta=\pi$, one can clearly see that each Ramsey sequence is an exact repetition of the previous sequence, since each sequence starts in state $|0\rangle$; due to the absence of correlations between successive measurements, the scaling corresponding to the standard quantum limit is expected. In contrast, in the coherent case, the information about $\sin (\theta /2)$ is stored in the amplitude of state $|2\rangle$ and then fed back into the Ramsey sequence at the next step. Overall, the evolution is unitary and therefore expected to be governed by Heisenberg scaling.

\section{Sources of errors} \label{sec_5}

In this section, we investigate the sensitivity of the protocol when subjected to sources of errors. In particular, the protocol's sensitivity to beam-splitter strength is important for assessing its effect on efficiency in the subsequent evolution. We also study the sensitivity of the protocol when arbitrary phases are introduced on the $B$ pulses, as well as the effect of having randomly placed $B$ pulses; i.e., some of the $B$ pulses which would normally occur in the Ramsey sequence, are switched off. The sensitivity of the protocol to the initial sample temperature, as well as the effects of decoherence via relaxation and of detuning, are also examined.

\subsection{ Effect of beam-splitter strength} \label{sec_5a}

\begin{figure}[h]
	\centering
\includegraphics[width=1.0\linewidth]{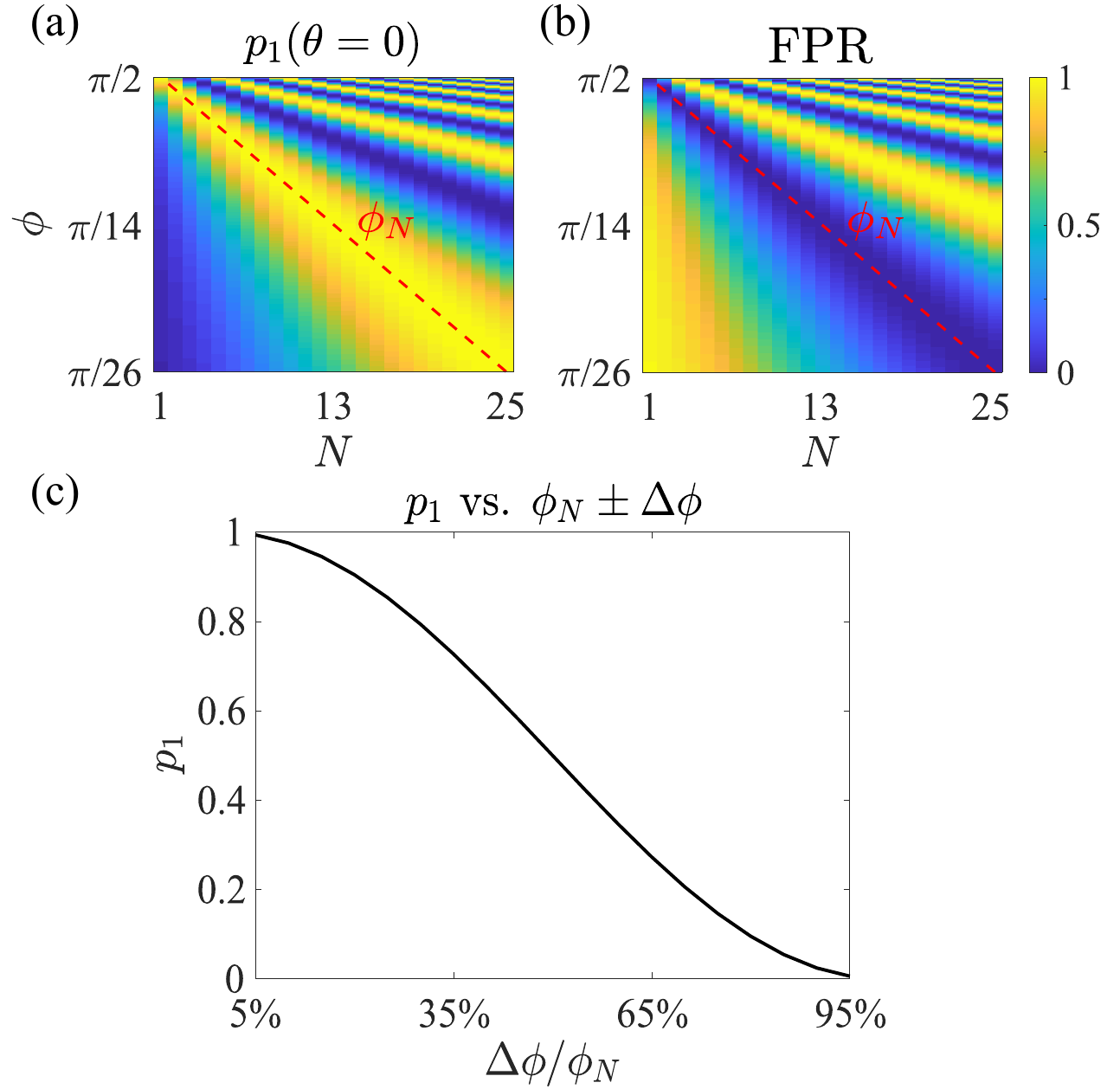}
	\caption{The first-excited state probability $p_{1}$ (a) and the false positive ratio $\rm FPR$ (b) as a function of  $N$ at various realizations of beam-splitter strength $\phi$ at $\theta_{j} = 0$, i.e., no $B$ pulses. For each $N$ the initial state is the ground state. (c) The sensitivity of $p_{1}$ as $\phi_N$ is varied by $\Delta \phi$. 
}
\label{fig-phi_optimal}
\end{figure}

We first consider the case $\theta_j = 0$, $\varphi_j =0$ and analyze the protocol with respect to $\phi$. The optimal choice of beam-splitter strength $\phi$ is  $\pi/(N+1)$~\cite{Kwiat_1995} for a protocol with $N$ $B$ pulses, which can be seen along the principal diagonal in Fig.~\ref{fig-phi_optimal}. This is the choice of beam-splitter strength such that only one of the two detectors in the projective protocol will click, and where there is a complete probability transfer from the initial state, i.e., the ground state, to the first excited state for our coherent protocol. 
Surface maps in Figs.~\ref{fig-phi_optimal}(a) and~\ref{fig-phi_optimal}(b) show the variation of the first-excited state probability $p_1$ and false positive ratio (FPR) as a function of  $N$ and beam-splitter angle $\phi$ for each $N \in [1,25]$. 
Curiously, there are other maxima in $p_{1}$ which can be seen in Fig.~\ref{fig-phi_optimal}. 
These maxima occur after every $2(N+1)$ beam-splitter unitaries for a given protocol with $N$ $B$ pulses. This results from the net rotation angle, i.e., $2(N+1)\phi=2\pi$, which after every $2(N+1)$ implementations brings the system back to the initial ground state with a phase of $e^{i \pi}$. From Fig.~\ref{fig-phi_optimal}(b), we see that FPR is high at low values of $\phi$. In other words, it is not always advantageous to have small $\phi$.

The sensitivity of the first-excited state probability $p_{1}$ to beam-splitter strength $\phi_N \pm \Delta \phi$ is shown in Fig.~\ref{fig-phi_optimal}(c) for strength $\theta = 0$. For $\theta=0$ and $p_2=0$, $p_1$ and $p_0$ are symmetric about $\phi_N$, i.e., $p_1(\phi_N +\Delta \phi) = p_1(\phi_N -\Delta \phi$). This behavior is independent of the chosen  $N$. We can also see that for low errors $\Delta \phi$, the first derivative of $p_1$ is zero, which makes $p_1$ sensitive only  to second-order in $\Delta \phi$  errors. For $\theta \neq 0$, we have $p_i(\phi_N + \Delta \phi) \neq p_{i}( \phi_N - \Delta \phi)$, $i \: \in \: \{0,1,2\}$, as $p_{i}$ is no longer independent of $N$.

For the case $\theta =\pi$, the projective protocol has a detection probability analytically expressed as
\begin{equation}
	p_{\rm det} = \left[\cos^2 \frac{\phi}{2} \right]^{(N+1)}
\end{equation}
and an absorption probability expressed by
\begin{equation}
	p_{\rm abs} = \sin^{2} \frac{\phi}{2} \sum_{j=1}^{N}\left[\cos^2 \frac{\phi}{2} \right]^{(j-1)} = \sin^{2} \frac{\phi}{2} \frac{1-\left[\cos^{2}\frac{\phi}{2}\right]^N}{1-\cos^{2}\frac{\phi}{2}}
	\;.
\end{equation}
These formulas can be obtained in a straightforward way from Eqs.\,(\ref{eq:det}) and\,(\ref{eq:abs}), and they coincide with those derived for Mach-Zehnder-based experiments in quantum optics \cite{Kwiat_1995, Kwiat_1999, Ma2014, Peise2015}. It is worth pausing and analyzing the meaning of these relations, as anticipated to some extent 
in the comment subsequent to Eqs.\,(\ref{eq:det}) and\,(\ref{eq:abs}). Starting in $|0\rangle$, the system remains in this state with probability $\cos^{2}(\phi/2)$ after the application of the first beam-splitter $S(\phi )$. If there is no absorption on state $|2\rangle$ after the $B$ pulse, it means that the second beam-splitter sees the system again in state $|0\rangle$. After $N+1$ applications  of the beam-splitter, the probability to find the system in the state $|0\rangle$ is $[\cos (\phi /2)] ^{2(N+1)}$.  In the case of absorption, $p_{\rm abs}$ is obtained by summing over probabilities of absorption at each application $j$ of the $B$ pulse, which are given by the probability $[\cos (\phi /2)] ^{2(j-1)}$ that the pulse was not absorbed in the previous $j-1$ steps, multiplied by the probability $\sin^{2}(\phi/2)$ that the system is in the state $|1\rangle$, from which absorption to state $|2\rangle$ is possible.

If $\phi = \phi_N = \pi/(N+1)$, the detection probability becomes 1 in the limit of large $N$, which is a manifestation of localization on the state $|0\rangle$ (suppressing the evolution in the rest of the Hilbert space) by the quantum Zeno effect. Indeed we have $\cos^2 (\phi_{N}/2) \approx 1 - \phi^2_{N}/4$ and by applying the binomial formula we obtain 
\begin{align}
p_{\rm det} \approx 1 - (N+1) \frac{\phi_N^2}{4} \approx 1 - \frac{\pi^2}{4N} 
\end{align}
and
\begin{align}
p_{\rm abs} \approx N \frac{\phi_{N}^2}{4} \approx \frac{\pi^2}{4N}\;,
\end{align}
which yields the efficiency $\eta \approx p_{\rm det}$. 
We can now see that, in contrast to the coherent case [see Eq.\,(\ref{eq:scale})], the scaling with $N$ of these probabilities is slower $\sim1/N$, indicative of the standard  quantum limit.

\begin{figure}[h]
\centering
\includegraphics[width=1.0\linewidth]{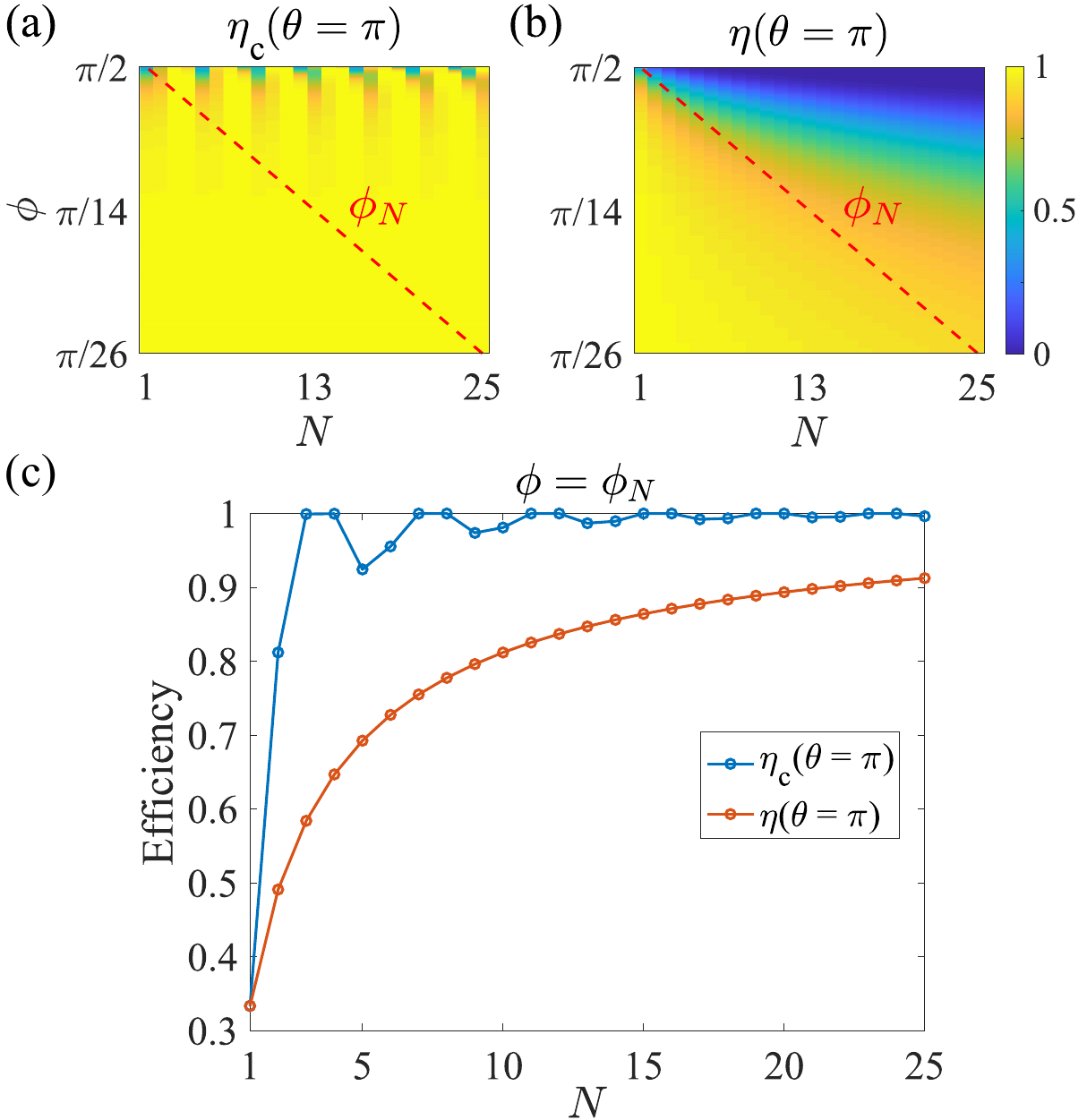}
\caption{
(a) The efficiency of the coherent protocol, $\eta_{\rm c}$, and (b) the projective protocol, $\eta$, as functions of $N$ for various realizations  of $\phi$ at $\theta_{j} = \pi$, $\varphi_{j} = \pi/2$. For each  $N$, the initial state is the ground state. (c) The efficiencies along the diagonal $\phi = \pi/(N+1)$ as a function of $N$. 
}\label{fig-efficiency_at_pi}
\end{figure}

In Figs.~\ref{fig-efficiency_at_pi}(a) and~\ref{fig-efficiency_at_pi}(b), the efficiencies resulting from the coherent protocol ($\eta_{\rm c}$) and projective protocol ($\eta$) are respectively plotted as functions of beam-splitter strength $\phi$ and $N$ (similar to Fig.~\ref{fig-phi_optimal}) at $B$-pulse strength $\theta=\pi$. 
In the upper triangular region where $\phi > \pi/(N+1)$ and near $\phi=\pi/2$, $\eta_{\rm c}$ is marginally higher, $\approx 1-2\%$, than the optimal value of $\eta_{\rm c}$ at $\phi=\pi/(N+1)$, only for a few values of $N$ (10, 14, 18, 21, 22, 25).
Lower values of $\eta$ in Fig.~\ref{fig-efficiency_at_pi}(b) for $\phi > \pi/(N+1)$ are mainly the result of a higher probability of occupation of the state $|1\rangle$, which further results in a higher probability for $p_{\rm abs}$. 

In the lower triangular section $\phi < \pi/(N+1)$, the efficiency reaches high values since the beam-splitter unitaries are not capable of transferring the ground state probability to the first excited state. This results in higher $p_0$ and hence an apparently higher $\eta_{\rm c}$ (as well as $\eta$), as seen in the lower triangular region of Figs.~\ref{fig-efficiency_at_pi}(a) and~\ref{fig-efficiency_at_pi}(b). However, one can see from Figs.~\ref{fig-phi_optimal}(a) and~\ref{fig-phi_optimal}(b) that the FPR increases to large values, making the protocol unusable.

Once again, we insist that $\phi=\pi/(N+1)$ is the optimal beam-splitter strength for a given $N$.
A neat comparison of the efficiencies $\eta$ (in red) and $\eta_{\rm c}$ (in blue) is presented in Fig.~\ref{fig-efficiency_at_pi}(c) for optimal values of the beam-splitter angle. Clearly, $\eta_{\rm c}$ already exceeds $0.95$ for $N>5$, while $\eta$ is below $0.9$ even for $N=25$, depicting a highly efficient coherent interaction-free measurement protocol as compared to that of the projective protocol.

An interesting situation is the case $\theta = 2\pi$. From Eq.\,(\ref{eq:Bmatrix}), we can see that the matrix at this choice of $\theta$ has $1$ followed by two $-1$'s on the diagonal, and zeros on the off-diagonal (thus making the phases $\varphi$ irrelevant). Classically, this is a $360^{o}$ rotation that should have no effect; yet quantum mechanically, due to the appearance of the minus signs, it has a dramatic effect. Indeed, $B (\theta = 2\pi)S(\phi ) |0\rangle = \cos\frac{\phi}{2}|0\rangle - \sin \frac{\phi}{2}|1\rangle$. We can see that the probability of absorption is zero! Further, after another application of $S(\phi)$, we obtain 
$S(\phi )B (\theta = 2\pi)S(\phi ) |0\rangle = |0\rangle $, therefore achieving a perfect interaction-free detection. Surprisingly, we have a situation where the efficiency of detecting a pulse that produces no absorption is maximal! Indeed, a detector based, say, on absorption of the pulse by a two-level system and the subsequent measurement of the excited-state probability would not be able to detect this pulse at all.

\subsection{$B$ pulses with a variable phase} \label{sec_5b}

Next, we consider the situation when both the $B$-pulse strength $\theta_{j}$ and phase $\varphi_{j}$ are non-zero. Here, we investigate the efficiency of the coherent protocol at different $N$ when subjected to various $\theta_{j}$ and $\varphi_{j}$, where $j \in [1,N]$. 

First, it is straightforward to verify that the results do not depend on the phase $\varphi_{j}$ in the projective case. This can be shown immediately by examining a sequence of Ramsey pulses with the measurement operators inserted after each $B$ pulse. The phase appears only on the state $|2\rangle$, and therefore it is eliminated when the non-absorptive result is obtained -- that is, from the application of Eq.\,(\ref{eq:Pnonabs}), $P_{\overline{\rm abs}} = |0\rangle \langle 0| + |1\rangle \langle 1|$. 

\begin{figure}[h]
\centering
\includegraphics[width=1.0\linewidth]{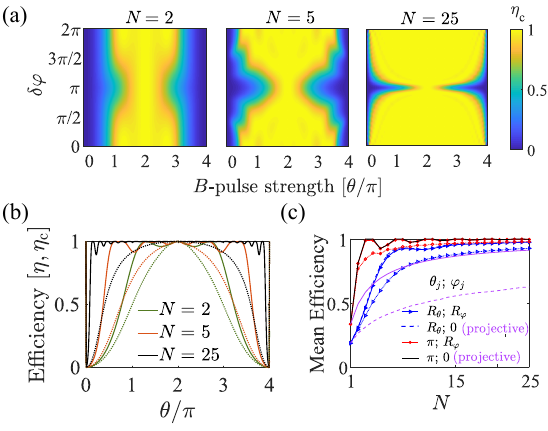}
\caption{
(a) The efficiency resulting from the coherent protocol $\eta_{\rm c}$ as a function of $B$-pulse strength $\theta \in [0,4\pi]$ and phase difference $\delta \varphi$ for $N = 2$, $5$ and $25$. (b) One-dimensional traces of $\eta_{\rm c}$ for $\delta \varphi = 0$ are plotted for $N=2$ (in green), $N=5$ (in red), and $N=25$ (in black) with solid lines, while dotted lines in respective colors represent the efficiency of the projective protocol ($\eta$) as a function of $\theta$ for the same values of $N$. 
(c) Variation of mean efficiency resulting from the coherent protocol ($\eta_{\rm c}^M$) is plotted as a function of $N$.
The solid black (magenta) curve corresponds to $\eta_{\rm c}$ ($\eta$) with $\theta_j=\pi$ and $\delta \varphi=0$, where $j \in [1,N]$. The red solid (dotted) curve with circular markers corresponds to constant $B$-pulse strengths $\theta_j = \pi$ and randomly chosen $B$-pulse phases $R_{\varphi}$ such that $\varphi_{j} \in [0, \pi/4]$ ($\varphi_{j} \in [0, \pi]$). The dashed blue (magenta) curve corresponds to the coherent (projective) mean efficiencies with randomly chosen $B$-pulse strengths $R_{\theta}$ such that $\theta_{j} \in [0, \pi]$, and with fixed $\varphi_{j}$, i.e., $\varphi_{j} = 0$. The solid (dotted) blue curve with triangular markers corresponds to $R_{\theta} \in [0,\pi]$ and $R_{\varphi} \in [0, \pi/4]$ ($R_{\varphi} \in [0, \pi]$). All of these plots are simulated with beam-splitters of strength $\phi = \pi/(N+1)$.}
\label{fig-varphi_vs_theta}
\end{figure}

In the coherent case, the efficiency depends on the phases only when the phase differences between consecutive $B$ pulses, $\delta \varphi \equiv \varphi_{j+1} - \varphi_{j}$, are nonzero. Fig.~\ref{fig-varphi_vs_theta}(a) shows $\eta_{\rm c}$ surface plots as a function of $\delta \varphi$ and $B$-pulse strength $\theta$ at $N = 2$, $N=5$, and $N=25$. It is clear from these surface maps that, for a wide range of $\delta \varphi$ values, we obtain wide plateaus of high efficiencies. It is also noteworthy that small values of $\delta \varphi$ do not cause any significant drop in the efficiency compared to $\delta \varphi=0$. The worst case corresponds to $\delta \varphi = \pi$, where these high-efficiency plateaus are significantly narrowed. 
The surface maps for $\eta_{\rm c}$ as a function of $\delta \varphi$ and $\theta_j=\theta$ are shown only for a few values of $N$; the behavior, however, is similar for other values of $N$.
The best case, i.e., a wide region of high $\eta_{\rm c}$ for arbitrary $N$, corresponds to $\delta \varphi=0$, which is plotted as solid lines in Fig.~\ref{fig-varphi_vs_theta}(b).

Moreover, Fig.~\ref{fig-varphi_vs_theta}(b) shows cross-sections of the aforementioned surface maps at $\delta \varphi =0$, as well as the projective efficiency $\eta$ as a function of $\theta$. Dotted lines in Fig.~\ref{fig-varphi_vs_theta}(b) are the corresponding efficiency plots for the projective case, with green, red, and black colors representing cases with $N=2,5$, and $25$, respectively. Clearly, as $N$ increases, higher efficiencies are attained over a broader range of $\theta$.

Next, we use the coherent efficiency $\eta_{\rm c}$ as a probe for performing detailed analysis of the protocol for arbitrary $B$ pulses with randomly chosen strengths and phases. Fig.~\ref{fig-varphi_vs_theta}(c) shows the mean efficiency ($\eta_{\rm c}^{M}$) vs $N$ for various choices of $\theta_j$ and $\varphi_j$ with $j \in [1,25]$. The mean efficiency for each $N$ is obtained from $10^4$ repetitions, each with a realization of $\theta_j$ and $\varphi_j$. 

The final probabilities, and hence $\eta_{\rm c}$, are independent of the $B$-pulse phase $\varphi$, if for a given $N$, all the $B$ pulses have the same phases, i.e., $\varphi_{j+1}=\varphi_j$, such that $\delta \varphi =\varphi_{j+1} - \varphi_j = 0$, where $j \in [1,N-1]$. This is verified numerically for various values of $N$ with arbitrarily chosen values of $\theta_j \in [0,\pi]$ and $\varphi_j=\varphi$, where $\varphi$ is chosen arbitrarily from the range $[0,\pi]$.
 
As a first check, we took $\theta_j = \pi$ and $\delta \varphi = 0$ and reproduced the blue curve representing $\eta_{\rm c}$ in Fig.~\ref{fig-efficiency_at_pi}(c) for different values of $\varphi$. This is represented as a solid black line in Fig.~\ref{fig-varphi_vs_theta}(c). Note that the solid magenta curve represents the efficiency of the projective case at $\theta_j = \pi$, i.e., identical to the red curve in Fig.~\ref{fig-efficiency_at_pi}(c). We also numerically verified that this property is extended for arbitrary values of $B$-pulse strengths $\theta_j$, which relaxes the specifications for the $B$ pulse. However, as previously seen in Fig.~\ref{fig-varphi_vs_theta}(a), relative differences between consecutive $B$-pulse phases ($\delta \varphi \neq 0$) can significantly alter the final probability profiles, and thus  $\eta_{\rm c}$.\newline \indent
Further, we study $\eta_{\rm c}^{M}$ when the phase is constant and the strengths are randomly varied such that $\theta_{j} \in [0,\pi]$ (denoted as $R_{\theta}$). Since a fixed phase does not affect the efficiency, we set $\varphi = 0$, as indicated in Fig.~\ref{fig-varphi_vs_theta}(c). The blue dashed line in the figure shows this case for the coherent protocol, and the magenta dashed line shows the corresponding mean efficiency $\eta^{M}$ for the projective case. \newline \indent
The red solid line with circular markers represents $\eta_{\rm c}^{M}$ when $\theta_{j} = \pi$ and the phases are randomly varied such that $\varphi_{j} \in [0,\pi/4]$ (labeled as $R_{\varphi}$). Clearly, $\eta_{\rm c}^{M}$ sits near the solid black line and is thus mostly insensitive to phase in this case. However, the efficiency is lower when the range of randomly varied phase is extended such that $R_{\varphi} \in [0,\pi]$. This is represented by the dotted red curve with circular markers in Fig.~\ref{fig-varphi_vs_theta}(c). Nevertheless, we conclude that small errors in the values of $\delta \varphi$ are tolerable without much compromise in the efficiency, which makes the coherent protocol robust to phase errors. \newline \indent
Further, there is a marked decrease in $\eta_{\rm c}^{M}$ when the $B$-pulse strengths are also random. In fact, the lowest mean efficiencies for the coherent protocol occur when $R_{\theta} \in [0, \pi]$ and $R_{\varphi} \in [0, \pi]$. This is shown as the blue dotted line with triangular markers. Only the projective cases shown in Fig.~\ref{fig-varphi_vs_theta}(c) tend to be lower than this case as $N$ becomes large. In particular, the mean projective efficiencies at $R_{\theta} \in [0, \pi]$ and $\varphi =0$ are consistently less than $\eta_{\rm c}^{M}(R_{\theta}, R_{\varphi})$ for $R_{\theta} \in [0, \pi]$ and $R_{\varphi} \in [0, \pi]$, and the projective efficiencies at $\theta_{j} = \pi$ and $\varphi_{j} = 0$ are also less than the lowest mean efficiencies of the coherent protocol after $N = 8$. Remarkably, this means that the coherent protocol is on average more efficient than the maximum efficiencies of the projective protocol, even when subjected to random $B$-pulse strengths and phases in the full range $[0, \pi]$. \newline \indent
The mean efficiencies when $R_{\theta} \in [0, \pi]$ and $R_{\varphi} \in [0, \pi/4]$ are represented as the solid blue curve with the triangular markers and are significantly larger than when $R_{\theta} \in [0, \pi]$ and $R_{\varphi} \in [0, \pi]$. Thus, as expected, the mean value $\eta_{\rm c}^{M}(R_{\theta}, R_{\varphi})$ lies close to the probabilities obtained with all $B$ pulses of strength $\pi$ and $\delta \varphi=0$ for large $N$~\cite{our_protocol}. 

\subsection{Interaction-free detection with randomly placed $B$ pulses} \label{sec_5c}

\begin{figure}[ht]
	\centering
	\includegraphics[width=0.85\linewidth,keepaspectratio=true]{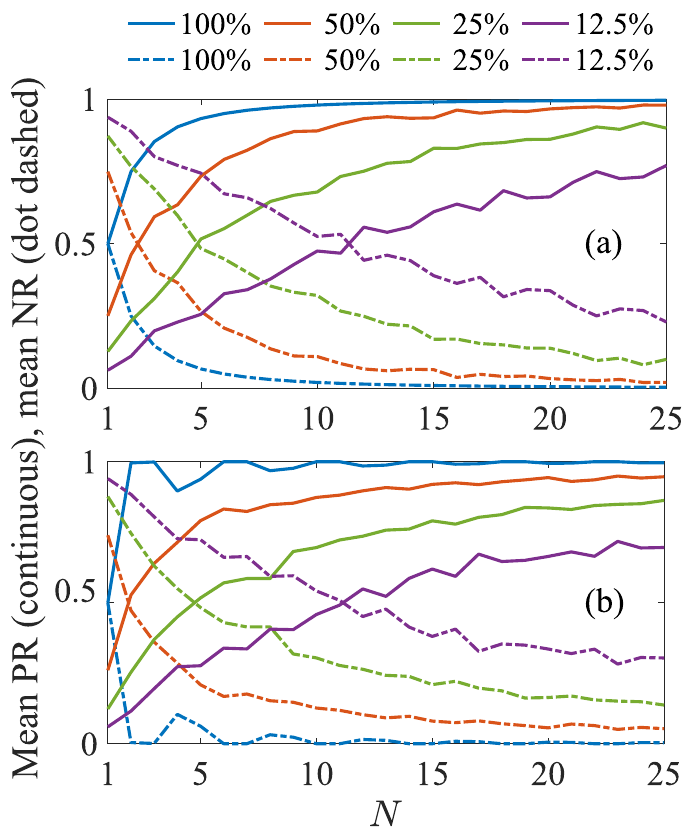}
	\caption{Positive (solid) and negative (dot-dashed) ratios are plotted for (a) projective and (b) coherent cases. On the legend (top of the figures) we show the percentages of $B$ pulses placed at random slots. The simulations are repeated 400 times.}
		\label{fig:fig5}
	\end{figure}

	In this section, we consider $N$ consecutive Ramsey sequences 
	with randomly placed $B$ pulses. In each Ramsey sequence,
	 the $B$-pulse slot can either have a $B$ pulse with $\theta=\pi$, i.e., maximum strength, or $\theta=0$ (no $B$ pulse). In other words, this situation corresponds to arbitrarily placing maximum-strength $B$ pulses in the $N$ Ramsey slots with a certain probability.
	Here, we consider four cases where each $B$-pulse slot can have a pulse with probabilities 1, 1/2, 1/4 and 1/8.
	Depending on the arrangements of the $B$ pulses in the full pulse sequence, the results can vary significantly.

	Suppose that out of $N$ $B$-pulse slots, n  have $B$ pulses with maximum strength while $N$-n are vacant. The number of combinations is $N!/(\rm{n}!$$(N-\rm{n})!)$, where $\rm{n}$ $\in [0,N]$. The total number of combinations can reach a maximum of $10^7$ at ${\rm n}=N/2$ and $N=25$.  
	
	Figure~\ref{fig:fig5}, shows the calculations of the positive ratio (PR) and negative ratio (NR) for (a) projective and (b) coherently interrogated detection schemes with different percentages of $B$ pulses. Each curve represents the average of PR or NR values obtained from 400 repetitions with random combinations of vacant and occupied slots for the $B$ pulses. 
	
	As shown in  Fig.~\ref{fig:fig5}, curves in blue correspond to a situation with all $B$ pulses of strength $\pi$, which means that there is a very large flux of microwave photons resonant with the $|1\rangle-|2\rangle$ transition. Due to this large flux, whenever level $|1\rangle$ acquires some population at the end of a beam-splitter operation, it is highly likely that our three-level system will transit to the second excited state. Despite the absorption of a fraction of photons, PR$(\theta)$ approaches 1, while the NR$(\theta)$ approaches 0. 
	
	It is interesting to note that as n decreases, the probability of photon absorption increases. This counterintuitive behavior is due to the abrupt and rapid decrease in the norm $p_0+p_1$. PR and NR are highly dependent on the specific combinations; therefore, it is more useful to look at their average behaviors. Consistent with these observations, it is also noteworthy that, as n decreases to $N/2$, events leading to the absorption of photons increase, which further increases for large $N$ values as n decreases to $N/4$ and $N/8$, respectively. 

\subsection{Initialization in thermal states} \label{sec_5d}

 The initial state of a real device is sometimes not perfectly thermalized to the ground state. For a real device such as the transmon, the initial state can have a rather high initial temperature, on the order of 50-100 mK \cite{Sultanov2021}.

	For a general three-level system in thermal equilibrium, the density matrix is of the form
	\begin{equation}
		\rho = p_{0}|0\rangle \langle 0| + p_{1}|1\rangle \langle 1| + p_{2}|2\rangle \langle 2|\;, 
		\label{eq:thermal_pops}
	\end{equation}
	where the probabilities are 
\begin{equation}
	p_{i} = \frac{1}{Z}e^{-\frac{E_{i}}{k_{B}T}}, \quad i \in \{0,1,2\}, \quad E_{0} = 0 \;,
\end{equation}
	and the canonical partition function is

\begin{equation*}
	Z = \sum_{i}\exp{[-E_{i}/k_{B}T]} = 1 + e^{-\hbar \omega_{01}/k_{B}{T}} + e^{-\hbar \omega_{02}/k_{B}{T}} \;.
\end{equation*}

\begin{figure}[h]
\centering
\includegraphics[width=1.0\linewidth]{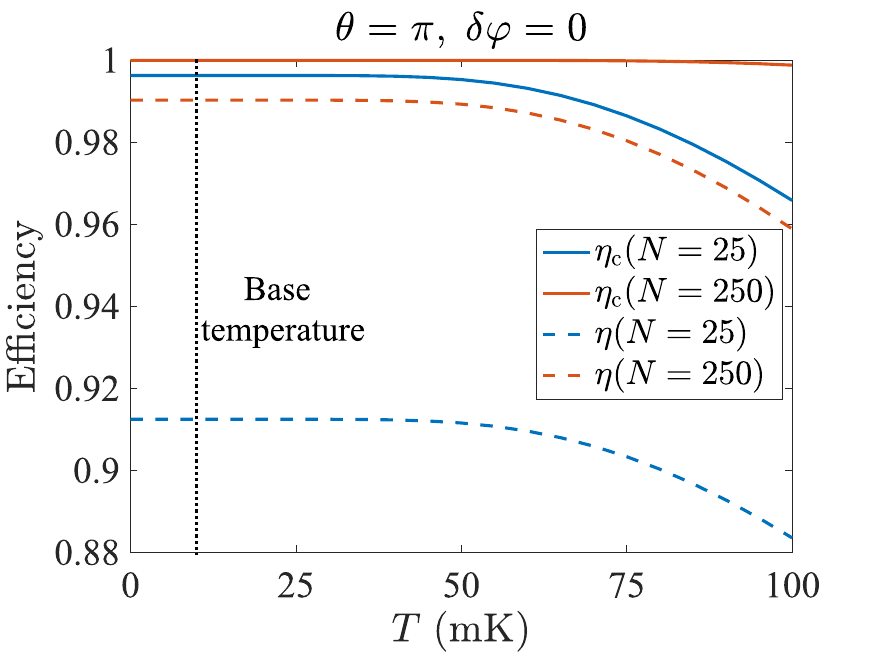}
\caption{Efficiencies for the two protocols vs initial temperatures $T \: \in \: [0, 100]$ mK, with each $B$-pulse strength at $\theta = \pi$ and no phase variation between consecutive $B$ pulses. Plots are shown for  $N$ = 25 and $N$ = 250 for each protocol. The typical base temperature of a dilution fridge (10 mK) is indicated with the black dotted line.
}
	\label{fig-initial-state}
\end{figure}

By populating our initial state in accordance with Eq.\,(\ref{eq:thermal_pops}) using qutrit transition frequencies $\omega_{01}/(2\pi) = 7.20$ GHz, $\omega_{12}/(2\pi) = 6.85$ GHz, and with initial temperatures $T \: \in \:[0, 100]$ mK, we see in Fig.~\ref{fig-initial-state} that the efficiencies are less sensitive at lower initial temperatures, and that the coherent protocol is overall more efficient than the projective case for a given $N$. In fact, at the modest  $N$ = 25, the efficiency of the coherent protocol is greater than the efficiency of the projective case at $N$ = 250.

	The dark count probabilities across this range of initial sample temperatures are determined by $p_{0}(\theta =0)$
at each temperature. These probabilities monotonically increase with temperature and are small across this range, less than $10^{-6}$ until 30 mK, and reach approximately $0.031$ at $100$ mK. These values are the same for both the coherent and projective protocols, as the dark count probabilities are necessarily computed at $\theta = 0$, and are independent of $N$ since we consistently choose $\phi = \pi/(N+1)$, i.e., $\phi_N$.

\subsection{Effects of decoherence} \label{sec_5e}

In real systems such as transmons, the action of the beam-splitters and the $B$ pulses 
is modified due to decoherence. To account for this effect, we consider a model where the first and second levels can relax to the ground and first excited state, respectively, with rates $\Gamma_{10}$ and $\Gamma_{21}$ \cite{Li2012,Guzik2011}.

The action of the beam-splitter on the density matrix is obtained from
\begin{equation}
	\dot{\rho} = -\frac{i}{\hbar}[H_{01,j}(t),\rho] + \sum_{l = 0,1; k= l+1}  \Gamma_{kl}D[\sigma_{lk}]\rho\;,
	\label{eq:Lind1}
\end{equation}
while for the $B$ pulse we have
\begin{equation}
	\dot{\rho} = -\frac{i}{\hbar}[H_{12,j}(t),\rho] + \sum_{l = 0,1; k= l+1}  \Gamma_{kl}D[\sigma_{lk}]\rho
	\label{eq:Lind2}
\end{equation}
with $H_{01,j}$ and $H_{12,j}$ as introduced in Sec.~\ref{sec_2},  and $D[L]\rho = L\rho L^{\dagger} - \frac{1}{2}\{L^{\dagger}L ,\rho \}$ defining the Lindblad super operator with jump operators $L$. Also note that, for the transmon, direct relaxation from the second excited state to the ground state is suppressed by selection rules.

\begin{figure}[h]
	 	\centering
	 	\includegraphics[width=0.95\linewidth,keepaspectratio=true]{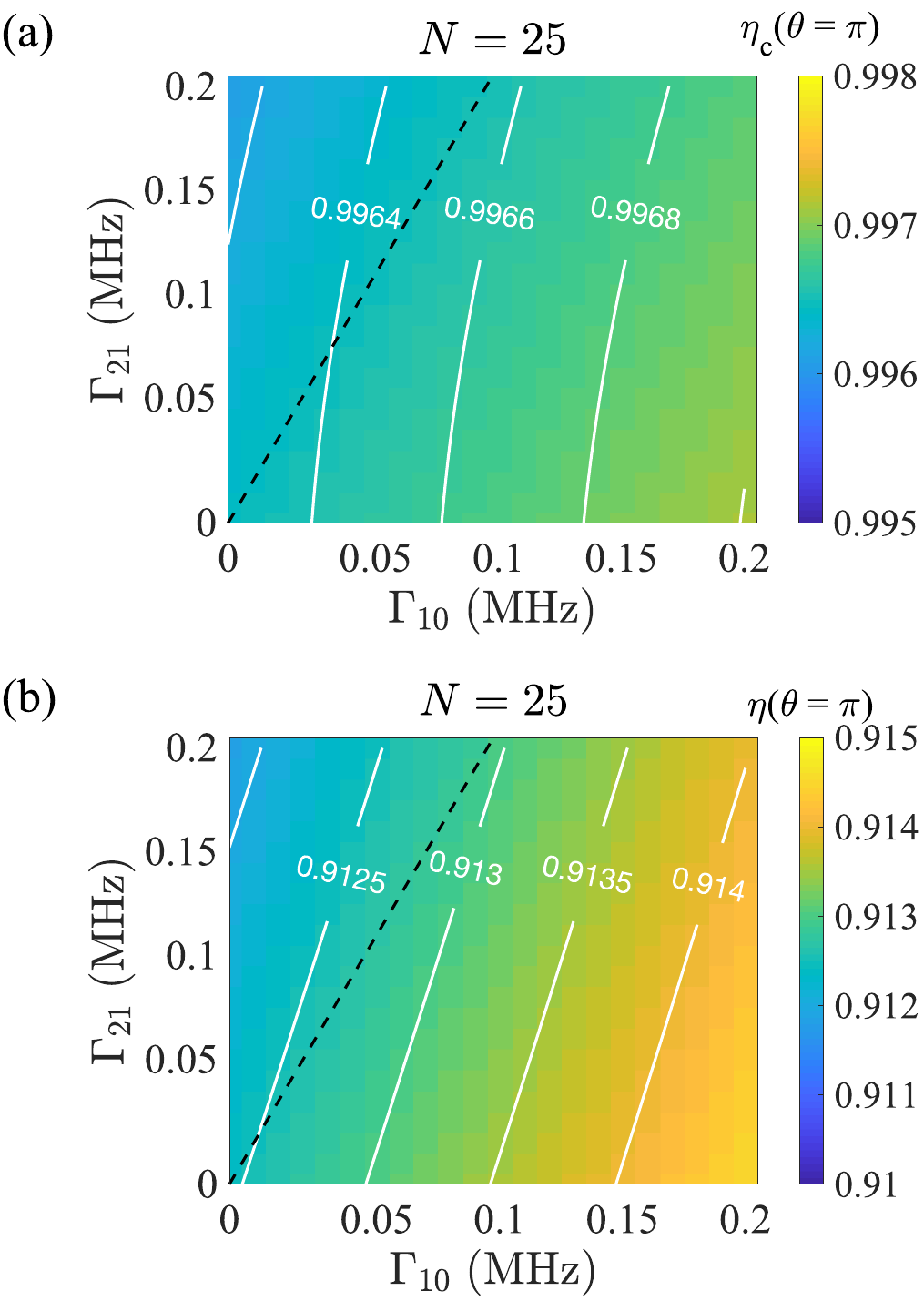}
	 	\caption{
	 		The efficiencies of the protocols, (a) $\eta_{\rm c}$ and (b) $\eta$, at $\theta = \pi$ and $N = 25$, are shown as functions of relaxation rates $\Gamma_{10}$ and $\Gamma_{21}$, both in MHz. The dashed line is along $\Gamma_{21} =  2\Gamma_{10}$, corresponding to a transmon.
		}
	 	\label{fig:deco}
	 \end{figure}

In Fig.~\ref{fig:deco}, we study the effect of various relaxation rates $\Gamma_{10}$, $\Gamma_{21} \in [0, 0.2]$ MHz on the efficiencies of the (a) coherent and (b) projective protocols at $\theta=\pi$. The dashed black line in both plots corresponds to the particular case of a transmon device, where these rates are related as $\Gamma_{21} =  2\Gamma_{10}$.

We see from Fig.~\ref{fig:deco} that $\eta_{\rm c}$ is consistently greater than $\eta$, where their mean values are approximately $0.9959$ and $0.9125$, respectively. We also note from the contour lines that the coherent case appears less sensitive to variation in $\Gamma_{10}$ compared to the projective protocol.
The dark count probabilities, i.e., FPRs, are also reasonably low, having a maximum value of $27.4\%$  for the worst-case scenario of a transmon with relatively large relaxation  $\Gamma_{10} = 0.2$ MHz. 

\begin{figure}[ht]
	\centering	\includegraphics[width=0.95\linewidth,keepaspectratio=true]{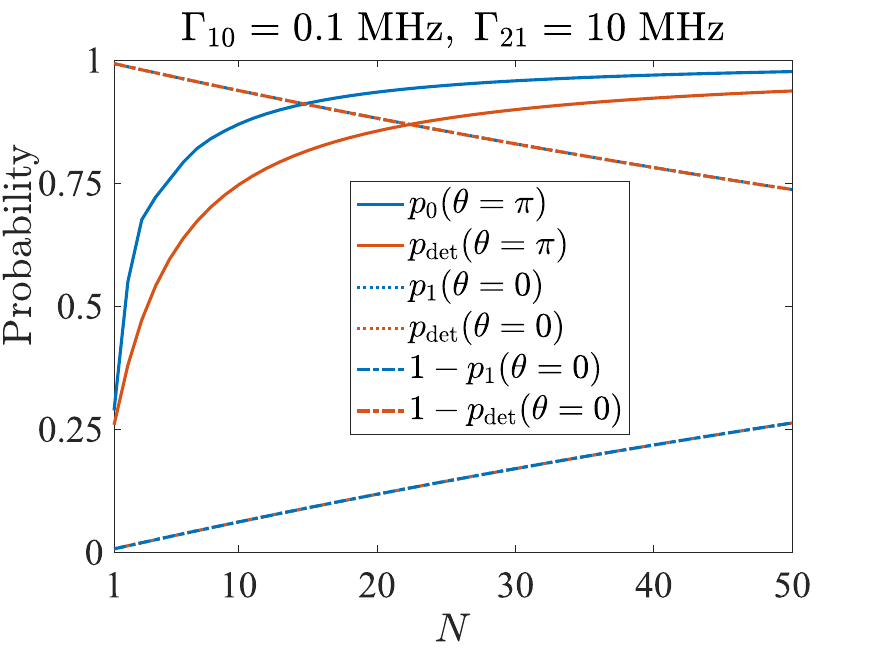}
	\caption{Probability vs $N \in [0, 50]$ for the coherent and projective protocols at relaxation rates $\Gamma_{10} = 0.1$ MHz, $\Gamma_{21} = 10$ MHz. In particular, $p_0$  and $p_{\rm det}$ are evaluated at $\theta = \pi$, and $p_{1}$ and $p_{\det}$, as well as their complements ($1- p_{1}$ and $1 - p_{\det}$) are evaluated at $\theta = 0$.}
	\label{fig:deco2}
\end{figure}

A remarkable feature of the coherent protocol is its robustness against decoherence acting on the $1-2$ subspace. One can see from Fig.~\ref{fig:deco}(a) that a change in $\Gamma_{21}$ produces a much smaller change in efficiency than a change in $\Gamma_{10}$ (equal-efficiency white lines are nearly vertical), an effect which is far less pronounced in the projective protocol. To illustrate this point, in Fig.~\ref{fig:deco2} we present $p_{0} (\theta =\pi)$ and $p_{1} (\theta =0)$ for $\Gamma_{10} = 0.1$ MHz and $\Gamma_{21} = 10$ MHz, with $N$ ranging from 1 to 50. Note that $\Gamma_{21}$ is 100 times larger than $\Gamma_{10}$ and yet the coherent protocol remains usable, with the limitation coming from the increase in the dark counts (FPR), $p_{0} (\theta =0) = 1- p_{1} (\theta =0)$, at large $N$. At $N = 50$, $p_{0}(\theta = \pi) = 0.977$ and $p_{\rm det}(\theta = \pi) = 0.937$ -- whereas, the dark count is $p_{0} (\theta =0) = 1- p_{1}(\theta = 0) = 0.263$. Clearly, the $p_{\rm det}$ of the projective protocol is more sensitive to $\Gamma_{21}$.  This difference is even more prominent at smaller values of $\theta$: numerically, we find that as $\theta$ decreases, both $p_{0} (\theta)$ and $p_{\rm det}(\theta)$ curves move to lower values maintaining a gap between them, with the latter approaching the $p_{\rm det}(\theta = 0) = p_{0}(\theta = 0)$ line more rapidly. The duration of each $B$ pulse is 112 ns, while that of beam-splitter pulses is 56 ns; therefore, 50 pulses take $8.5~\mu$s, much longer than the relaxation time $\Gamma_{21}^{-1} = 100$ ns of the state $|2\rangle$. This can be understood by the fact that the $B$ pulse and the relaxation act jointly as a disturbance of the interferometric pattern. It also shows that, in order to apply our protocol, one only needs a good two-level system: even if the third state is affected by large decoherence, the protocol will still work. 

Finally, even if the FPR is affected in a relatively stronger way by the relaxation in the $0-1$ subspace due to the increase in the dark count probability $p_0(\theta=0)$, this detrimental effect is still slightly weaker than what one would expect from a naive estimation of probabilities decaying exponentially with a rate $\Gamma_{10}$ during the total duration of the protocol.

\subsection{Detuned $B$ pulses} \label{sec_5f}

\begin{figure}[h]
	\centering
	\includegraphics[width=1\linewidth]{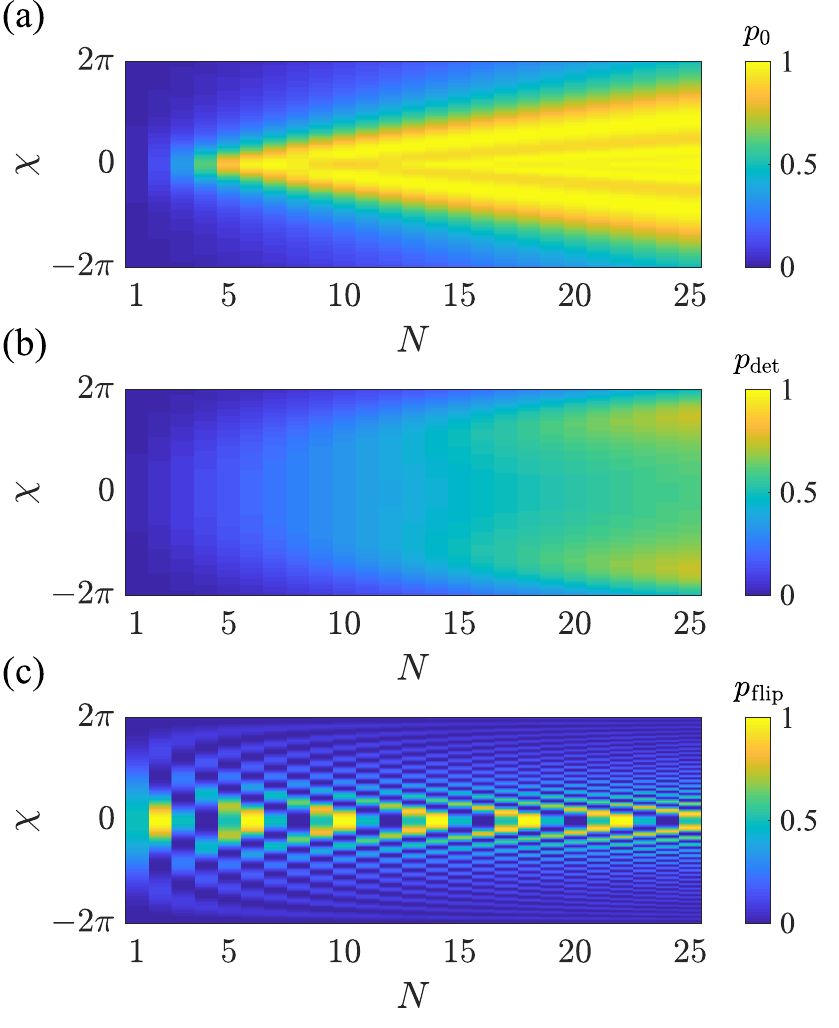}
	\caption{The success probabilities as functions of detuning $\chi$ and $N$ Ramsey sequences for (a) the coherent case and (b) the projective case at $\theta = \pi/2$. Panel (c) shows the probability of flipping, $p_{\rm flip}$, in the corresponding two-level system at $\theta = \pi/2$, as a function of $\chi$ and $N$, when the evolution is solely governed by $\mathcal{B}$ acting on the state $(0,~1)^{\rm T}$ $N$ times.} 
	\label{fig-detuning}
\end{figure}

We now examine the effect of a detuning $\delta$ of the $B$ pulse with respect to the second transition. For simplicity, we consider identical pulses, implemented by the Hamiltonian $H_{12}(t) = \hbar[\Omega_{12}(t)\exp(-i\varphi)/2]|1\rangle \langle 2| + \textrm{h.c.} - \hbar \delta |2\rangle\langle 2|$. With the usual notation $\theta = \int_{-\infty}^{\infty}\Omega_{12}(t) dt$ and $\chi = \delta \tau$, where $\tau$ is the duration of the $B$ pulse, we find that $B(\theta,\varphi, \chi)$ takes the form

\begin{equation}
B(\theta,\varphi, \chi) = 	
	\left( \begin{array}{cc} 1 & O_{1\times2} \\ O_{2\times1} & e^{i \chi/2} \mathcal{B} \end{array}
	\right),
\end{equation}
where again $O_{n1 \times n2}$ is the null matrix of dimension $n_1 \times n_2$ and the submatrix $\mathcal{B}$ has elements 
\begin{equation}
\mathcal{B}_{11} = \mathcal{B}_{22}^{*} = \cos\frac{\sqrt{\theta^2 + \chi^2}}{2} - \frac{i \chi}{\sqrt{\theta^2 + \chi^2}}\sin\frac{\sqrt{\theta^2 + \chi^2}}{2},
\end{equation}
and
\begin{equation} 
\mathcal{B}_{12} = -\mathcal{B}_{12}^{*} = - \frac{i \theta e^{-i\varphi}}{\sqrt{\theta^2 + \chi^2}} \sin \frac{\sqrt{\theta^2 + \chi^2}}{2}.
\end{equation}
In Figs.~\ref{fig-detuning}(a) and~\ref{fig-detuning}(b), we present the results of simulating the protocol up to $N=25$ for $\theta = \pi /2$. 
Remarkably, for the coherent case, as $N$ gets larger, the $B$ pulse can be detected even for relatively large values of $\chi$. In other words, the small effect on the interference pattern at small values of $N$ gets amplified at larger $N$. In contrast, this effect is not so prominent for the projective case. The detection bandwidth of $p_{0}$ appears to linearly increase symmetrically about $\chi = 0$ producing a fan-out structure, whereas $p_{\rm det}$ has less defined features and lower values. 

To show how dramatically different this situation is from the two-level case, consider what would happen if we aim to detect the pulse by measuring the off-resonant Rabi oscillation produced by a pulse $\mathcal{B}$ acting $N$ times on an initial state with maximum population on one of the levels. Fig.~\ref{fig-detuning}(c) shows the population on the other level, $p_{\rm flip}$, which can be used as a detection signal and is explicitly
\begin{equation}
	p_{\rm flip} = \frac{\theta^2}{\theta^2 + \chi^2}\sin^2 \frac{N\sqrt{\theta^2 + \chi^2}}{2}\;.
\end{equation}
We can see that the detection bandwidth does not increase with $N$.

\section{Conclusions} \label{sec_6}

We have investigated a protocol for interaction-free measurements in a three-level system that uses coherent unitary evolution instead of projective measurements. We found that the coherent scheme is generally more efficient than the projective protocol, and we derived asymptotic analytical results that demonstrate conclusively the existence of this enhancement. When considering the large $N$ limit, we determined the minimum value of $B$-pulse strength which yields optimal success probability and efficiency to be approximately four times the beam splitter strength. From the analysis of Fisher information, we found that for weak $B$ pulses our coherent interaction-free detection scheme reaches the Heisenberg limit while the projective scheme may only reach the standard quantum limit. We also numerically explored the sensitivity of our coherent interaction-free detection scheme under various imperfections and realistic conditions and compared it with the projective one. We find that the coherent protocol remains robust under experimentally relevant variations in beam-splitter strengths, temperature, decoherence, and detuning errors. Our results open up a new route towards quantum advantage by proposing a task that cannot be achieved classically and by using coherence as a quantum resource to achieve it efficiently.

\acknowledgments  
 We acknowledge financial support from the Finnish Centre of Excellence in Quantum Technology QTF (Projects No. 312296, No. 336810, No. 352925) of the Academy of Finland and from Business Finland QuTI (Decision No. 41419/31/2020).
 
\appendix

\section{Details about the derivation of analytical results in the large-$N$ limit} \label{Appendix_A}

Here we give more details about the diagonalization of the operator $S(\phi_N)B(\theta)$. We expand $\mathrm{det}[S(\phi_N)B(\theta)-\lambda \mathbb{I}_3]$ in powers of $\phi_{N}/2$ and retain terms up to second order, obtaining
\begin{eqnarray}
	&&\mathrm{det}[S(\phi_N)B(\theta)-\lambda \mathbb{I}_3] \approx (1-\lambda )	[\lambda^2 -2\lambda \cos(\theta /2) +1 \nonumber \\
	&& - \frac{1}{2}(\phi_N/2)^2(1-\lambda \cos (\theta/2))]  -\frac{1}{2}(\phi_N/2)^2[\lambda^2 -2\lambda \cos(\theta /2) \nonumber\\
	&& + 1] + (\phi_{N}/2)^2[1- \lambda\cos(\theta/2)]. \nonumber
\end{eqnarray}
Next, we neglect terms of the type $(\phi_N/2)^2\cos(\theta/2)$. Note that this is a better approximation than just working around $\theta \approx \pi$, allowing us to retain $\cos(\theta/2)$ in the expression above whenever it does not get multiplied by the small factor $(\phi_N/2)^2$.
In this approximation, after some algebra, we obtain the eigenvalues $\lambda_0 = 1$, $\lambda_{\pm} = \pm \exp (\pm i \theta /2)$, with corresponding eigenvectors
\medskip
$|v_0 \rangle= \left( \begin{array}{c}
	1 \\ \tan(\phi_N/4) \\ \tan(\phi_N/4)\, \cot(\theta/4)
\end{array} \right)$,                    
$|v_\pm \rangle = \left( \begin{array}{c}
	\sin (\phi_N/2) \\ a \mp ib \\ \mp ia  - b 
\end{array} \right)$, \medskip where $a = \cos (\theta /2) \cos (\phi_{N}/2) -1$ and $b = \sin (\theta /2) \cos (\phi_N /2)$. 

Note that $|v_{\pm}\rangle = |v_{\mp}^{*}\rangle$.  
We get
\begin{displaymath}
	\left[ S (\phi_N) B(\theta) \right]^{N+1} = M \cdot \left( \begin{array}{ccc}
		1 & 0 & 0 \\ 
		0 & e^{-i (N+1)\theta/2} & 0 \\ 
		0 & 0 & e^{i (N+1)\theta/2}
	\end{array} \right) \cdot M^{-1},
\end{displaymath}
where
\begin{displaymath}
	M = \left( \begin{array}{ccc}
		1 & \sin (\phi_N /2) & \sin (\phi_N /2) \\ 
		\tan(\phi_N/4) & a + i b & a - ib \\ 
		\tan(\phi_N/4) \, \cot(\theta/4) & ia - b  & -i a -b 
	\end{array} \right) .
\end{displaymath}
Using the above expressions, we obtain an approximate final state ($c_0 |0\rangle + c_1 |1\rangle + c_2 |2\rangle $) of the three-level system,

\begin{eqnarray}
	c_0 &=& \frac{1}{\mathcal{N}} \left[
	(a^2 + b^2) \sin \frac{\theta}{4}
	+ \tan \frac{\phi_N}{4}\, \sin \frac{\phi_N}{2} \right. \nonumber \\
	&& \left. \times 
	\left( a \sin \frac{(2N + 1)\theta}{4}
	+ b \cos \frac{(2N + 1)\theta}{4} \right)
	\right],
	\label{Eq:c0} \\[6pt]
	c_1 &=& \frac{2 (a^2 + b^2)}{\mathcal{N}}
	\tan \frac{\phi_N}{4}
	\sin \frac{(N + 1)\theta}{4}
	\cos \frac{N\theta}{4},
	\label{Eq:c1} \\[6pt]
	c_2 &=& \frac{2 (a^2 + b^2)}{\mathcal{N}}
	\tan \frac{\phi_N}{4}
	\sin \frac{(N + 1)\theta}{4}
	\sin \frac{N\theta}{4},
	\label{Eq:c2}
\end{eqnarray}
where \[ \mathcal{N} = (a^2 + b^2)\sin\frac{\theta}{4}
- \tan\frac{\phi_N}{4}\,\sin\frac{\phi_N}{2}
\bigl[ a\sin\frac{\theta}{4} - b\cos\frac{\theta}{4} \bigr].
\]

Note that here we have approximated $\sin(\phi_N /2) \approx \phi_N /2$ and $\cos(\phi_N /2) \approx 1 - (\phi_N /2)^2/2$ when calculating the eigenvalues, but we have chosen to keep the full trigonometric expressions 
 for the eigenvectors (and subsequently in the expressions for the matrix $M$ and for the coefficients $c_{0}$, $c_{1}$, and $c_{2}$). Compared to the case in the main text, where everything has been Taylor expanded in $\phi_N/2$, this leads to a slightly better approximation, especially at low values of $\theta$, albeit at the expense of more complicated analytical expressions.

The probability amplitudes Eqs.\,(\ref{Eq:c0})--(\ref{Eq:c2}) agree well with the probability amplitudes derived in the main text, i.e., Eqs.\,(\ref{eq:c0approx})--(\ref{eq:c2approx}). For instance, at $\theta = \pi$ and $N = 5$, these are $c_0  = 0.933$ $(0.931)$, $c_1 = 0.254$ $(0.262)$, and $c_2 = 0.254$ $(0.262)$, where the values in parenthesis correspond to Eqs.\,(\ref{eq:c0approx})--(\ref{eq:c2approx}), respectively. At $\theta = \pi$ and $N = 25$, these are $c_0$ = 0.996 (0.996), $c_{1} $= 0.0603 (0.0604), and $c_{2}$ = 0.0603 (0.0604).

The coherent detection efficiency of the protocol is given by
\begin{equation}
	\eta_{\rm c}=\frac{p_0}{p_0 + p_2} = \frac{|c_0|^2}{|c_0|^2 + |c_2|^2},
\end{equation}
which, at $\theta=\pi$, is approximated as

\begin{equation}
\eta_{\rm c} (\theta=\pi) = 1- \frac{\phi_{N}^2}{16} \left[1-\sqrt{2} \cos \left( \frac{N\pi}{2} + \frac{\pi}{4}\right) \right]^2\;. \label{eq:scale}
\end{equation}
This is in agreement with the results in the main text.

\section{Recursion relations} \label{Appendix_B}

In this appendix, we detail how the state at each step is related recursively to the previous Ramsey sequence. 
Let us consider $N$ Ramsey sequences with the beam-splitter strength $\phi$, $B$-pulse strength $\theta_{j}$, and phase $\varphi_{j}$, where $j \in [1,N]$.

For the projective case, the state after the application of $j$ sequences is 
 $S(\phi) \prod_{i=1}^{j}\left[P_{\overline{\rm abs}}B(\theta_{i}, \varphi_{i})S(\phi)\right]|0\rangle$. 
Let us denote this state generically by $c_{j,0}|0\rangle + c_{j,1}|1\rangle + c_{j,2}|2\rangle$.
Therefore the (unnormalized) probability amplitudes are recursively related to the subsequent values at $j+1$ as follows:
\begin{eqnarray}
c_{j+1,0} &=& \cos \frac{\phi}{2}c_{j,0} - \sin \frac{\phi}{2} \cos \frac{\theta_{j+1}}{2} c_{j,1}\;, \label{eq:recursion1} \\
c_{j+1,1} &=& \sin \frac{\phi}{2} c_{j,0} + \cos \frac{\phi}{2} \cos \frac{\theta_{j+1}}{2} c_{j,1}\;, \label{eq:recursion2} \\
c_{j+1,2} &=& c_{j,2} = 0\;.
\label{eq:recursion3}
\end{eqnarray}

For the coherent case, the state after applying $j$ Ramsey sequences is given by $S(\phi) \prod_{i=1}^{j}\left[B(\theta_{i}, \varphi_{i})S(\phi)\right]|0\rangle$.  
Let us similarly denote this wavefunction as $c_{j,0}|0\rangle + c_{j,1}|1\rangle + c_{j,2}|2\rangle$. Then the probability amplitudes $c_{j,0}$, $c_{j,1}$, and $c_{j,2}$
satisfy the following recursion relations
\begin{eqnarray}
c_{j+1,0} &=& \cos \frac{\phi}{2}c_{j,0} 
 - \sin \frac{\phi}{2}\cos \frac{\theta_{j+1}}{2}c_{j,1} \label{eq:recursion_1} \\  
 & &+ ie^{-i\varphi_{j+1}}\sin \frac{\phi}{2} \sin \frac{\theta_{j+1}}{2}c_{j,2},  \nonumber \\ 
c_{j+1,1} &=& \sin \frac{\phi}{2}c_{j,0}  + \cos \frac{\phi}{2}\cos \frac{\theta_{j+1}}{2}c_{j,1} \label{eq:recursion_2} \\ 
& &- ie^{-i\varphi_{j+1}}\cos \frac{\phi}{2} \sin \frac{\theta_{j+1}}{2}c_{j,2}, \nonumber \\ 
c_{j+1,2} &=& -ie^{i\varphi_{j+1}}\sin \frac{\theta_{j+1}}{2}c_{j,1} + \cos \frac{\theta_{j+1}}{2}c_{j,2}\;. 
\label{eq:recursion_3}
\end{eqnarray}

With these notations, we also have, at the end of the sequences, that $c_{0} \equiv c_{N,0}$,  $c_{1} \equiv c_{N,1}$, $c_{2} \equiv c_{N,2}$, making the connection with the previous usage of coefficients $c_0$, $c_1$, and $c_2$ for the coherent case. 

In both cases, we can now see the mechanism by which the probability corresponding  to the ground state increases under successive sequences. Indeed, if at some $j$ we have $|c_{j,0}| \approx 1$ (and consequently $|c_{j,1}| \ll 1$, $|c_{j,2}| \ll 1$), in the next step, $c_{j+1,1}$ will acquire a contribution $\sin (\phi /2) c_{j,0}$, which is very small since $\phi \ll 1$. We will also acquire a contribution 
from the very small previous values $c_{j,1}$ and $c_{j,2}$ (in the coherent case). In contrast, $c_{j+1,0}$ acquires a contribution $\cos (\phi /2)c_{j,0} \approx 1$, therefore remaining the dominant probability amplitude. At the end of the sequence and for $N \gg 1$, the state will be $|0\rangle$, in agreement with the observations from Sec.~\ref{sec_3a} related to Eq.\,(\ref{eq:matrx}).

Numerical simulations of the probabilities of success [$p_0(j,N)$, $p_{\rm det}(j,N)$] and of absorption [$p_2(j,N)$, $p_{\rm abs}(j,N)$], shown in Fig.~\ref{fig:fig6}, directly correspond to the absolute squares of the corresponding complex coefficients discussed above.


\begin{thebibliography}{33}%
\makeatletter
\providecommand \@ifxundefined [1]{%
 \@ifx{#1\undefined}
}%
\providecommand \@ifnum [1]{%
 \ifnum #1\expandafter \@firstoftwo
 \else \expandafter \@secondoftwo
 \fi
}%
\providecommand \@ifx [1]{%
 \ifx #1\expandafter \@firstoftwo
 \else \expandafter \@secondoftwo
 \fi
}%
\providecommand \natexlab [1]{#1}%
\providecommand \enquote  [1]{``#1''}%
\providecommand \bibnamefont  [1]{#1}%
\providecommand \bibfnamefont [1]{#1}%
\providecommand \citenamefont [1]{#1}%
\providecommand \href@noop [0]{\@secondoftwo}%
\providecommand \href [0]{\begingroup \@sanitize@url \@href}%
\providecommand \@href[1]{\@@startlink{#1}\@@href}%
\providecommand \@@href[1]{\endgroup#1\@@endlink}%
\providecommand \@sanitize@url [0]{\catcode `\\12\catcode `\$12\catcode
  `\&12\catcode `\#12\catcode `\^12\catcode `\_12\catcode `\%12\relax}%
\providecommand \@@startlink[1]{}%
\providecommand \@@endlink[0]{}%
\providecommand \url  [0]{\begingroup\@sanitize@url \@url }%
\providecommand \@url [1]{\endgroup\@href {#1}{\urlprefix }}%
\providecommand \urlprefix  [0]{URL }%
\providecommand \Eprint [0]{\href }%
\providecommand \doibase [0]{http://dx.doi.org/}%
\providecommand \selectlanguage [0]{\@gobble}%
\providecommand \bibinfo  [0]{\@secondoftwo}%
\providecommand \bibfield  [0]{\@secondoftwo}%
\providecommand \translation [1]{[#1]}%
\providecommand \BibitemOpen [0]{}%
\providecommand \bibitemStop [0]{}%
\providecommand \bibitemNoStop [0]{.\EOS\space}%
\providecommand \EOS [0]{\spacefactor3000\relax}%
\providecommand \BibitemShut  [1]{\csname bibitem#1\endcsname}%
\let\auto@bib@innerbib\@empty
\bibitem [{\citenamefont {Elitzur}\ and\ \citenamefont
		{Vaidman}(1993)}]{Elitzur_1993}%
	\BibitemOpen
	\bibfield  {author} {\bibinfo {author} {\bibfnamefont {A.~C.}\ \bibnamefont
			{Elitzur}}\ and\ \bibinfo {author} {\bibfnamefont {L.}~\bibnamefont
			{Vaidman}},\ }\bibfield  {title} {\bibinfo {title} {Quantum mechanical
			interaction-free measurements},\ }\href {https://doi.org/10.1007/bf00736012}
	{\bibfield  {journal} {\bibinfo  {journal} {Foundations of Physics}\ }\textbf
		{\bibinfo {volume} {23}},\ \bibinfo {pages} {987–997} (\bibinfo {year}
		{1993})}\BibitemShut {NoStop}%
\bibitem [{\citenamefont {Renninger}(1953)}]{Renninger}%
	\BibitemOpen
	\bibfield  {author} {\bibinfo {author} {\bibfnamefont {M.}~\bibnamefont
			{Renninger}},\ }\bibfield  {title} {\bibinfo {title} {Zum wellen-korpuskel-dualismus},\\ \ }\href{https://doi.org/10.1007/BF01325679} {\bibfield  {journal} {\bibinfo
			{journal} {Zeitschrift f{\"u}r Physik}\ }\textbf {\bibinfo {volume}
			{136}},\ \bibinfo {pages} {251} (\bibinfo {year} {1953})}\BibitemShut
	{NoStop}%
\bibitem [{\citenamefont {Dicke}(1981)}]{Dicke}%
  \BibitemOpen
  \bibfield  {author} {\bibinfo {author} {\bibfnamefont {R.~H.}\ \bibnamefont
  {Dicke}},\ }\bibfield  {title} {\bibinfo {title} {Interaction-free
  quantum measurements: A paradox?},\ }\href {\doibase 10.1119/1.12592}
  {\bibfield  {journal} {\bibinfo  {journal} {American Journal of Physics}\
  }\textbf {\bibinfo {volume} {49}},\ \bibinfo {pages} {925--930} (\bibinfo
  {year} {1981})}\BibitemShut {NoStop}%
\bibitem [{\citenamefont {Peres}(1980)}]{peres-ajp-1980}%
  \BibitemOpen
  \bibfield  {author} {\bibinfo {author} {\bibfnamefont {A.}~\bibnamefont
  {Peres}},\ }\bibfield  {title} {\bibinfo {title} {Zeno paradox in
  quantum theory},\ }\href {\doibase 10.1119/1.12204} {\bibfield  {journal}
  {\bibinfo  {journal} {Am. J. Phys.}\ }\textbf {\bibinfo {volume} {48}},\
  \bibinfo {pages} {931} (\bibinfo {year} {1980})}\BibitemShut {NoStop}%
\bibitem [{\citenamefont {Kwiat}\ \emph {et~al.}(1995)\citenamefont {Kwiat},
		\citenamefont {Weinfurter}, \citenamefont {Herzog}, \citenamefont
		{Zeilinger},\ and\ \citenamefont {Kasevich}}]{Kwiat_1995}%
	\BibitemOpen
	\bibfield  {author} {\bibinfo {author} {\bibfnamefont {P.}~\bibnamefont
			{Kwiat}}, \bibinfo {author} {\bibfnamefont {H.}~\bibnamefont {Weinfurter}},
		\bibinfo {author} {\bibfnamefont {T.}~\bibnamefont {Herzog}}, \bibinfo
		{author} {\bibfnamefont {A.}~\bibnamefont {Zeilinger}},\ and\ \bibinfo
		{author} {\bibfnamefont {M.~A.}\ \bibnamefont {Kasevich}},\ }\bibfield
	{title} {\bibinfo {title} {Interaction-free measurement},\ }\href
	{https://doi.org/10.1103/PhysRevLett.74.4763} {\bibfield  {journal} {\bibinfo
			{journal} {Phys. Rev. Lett.}\ }\textbf {\bibinfo {volume} {74}},\ \bibinfo
		{pages} {4763} (\bibinfo {year} {1995})}\BibitemShut {NoStop}%
\bibitem [{\citenamefont {Kwiat}\ \emph {et~al.}(1999)\citenamefont {Kwiat},
  \citenamefont {White}, \citenamefont {Mitchell}, \citenamefont {Nairz},
  \citenamefont {Weihs}, \citenamefont {Weinfurter},\ and\ \citenamefont
  {Zeilinger}}]{Kwiat_1999}%
  \BibitemOpen
  \bibfield  {author} {\bibinfo {author} {\bibfnamefont {P.~G.}\ \bibnamefont
  {Kwiat}}, \bibinfo {author} {\bibfnamefont {A.~G.}\ \bibnamefont {White}},
  \bibinfo {author} {\bibfnamefont {J.~R.}\ \bibnamefont {Mitchell}}, \bibinfo
  {author} {\bibfnamefont {O.}~\bibnamefont {Nairz}}, \bibinfo {author}
  {\bibfnamefont {G.}~\bibnamefont {Weihs}}, \bibinfo {author} {\bibfnamefont
  {H.}~\bibnamefont {Weinfurter}}, and \bibinfo {author} {\bibfnamefont
  {A.}~\bibnamefont {Zeilinger}},\ }\bibfield  {title} {\bibinfo
  {title} {High-efficiency quantum interrogation measurements via the quantum
  Zeno effect},\ }\href {\doibase 10.1103/physrevlett.83.4725} {\bibfield
  {journal} {\bibinfo  {journal} {Physical Review Letters}\ }\textbf {\bibinfo
  {volume} {83}},\ \bibinfo {pages} {4725–4728} (\bibinfo {year}
  {1999})}\BibitemShut {NoStop}%
\bibitem [{\citenamefont {Ma}\ \emph {et~al.}(2014)\citenamefont {Ma},
		\citenamefont {Guo}, \citenamefont {Schuck}, \citenamefont {Fong},
		\citenamefont {Jiang},\ and\ \citenamefont {Tang}}]{Ma2014}%
	\BibitemOpen
	\bibfield  {author} {\bibinfo {author} {\bibfnamefont {X.-s.}\ \bibnamefont
			{Ma}}, \bibinfo {author} {\bibfnamefont {X.}~\bibnamefont {Guo}}, \bibinfo
		{author} {\bibfnamefont {C.}~\bibnamefont {Schuck}}, \bibinfo {author}
		{\bibfnamefont {K.~Y.}\ \bibnamefont {Fong}}, \bibinfo {author}
		{\bibfnamefont {L.}~\bibnamefont {Jiang}},\ and\ \bibinfo {author}
		{\bibfnamefont {H.~X.}\ \bibnamefont {Tang}},\ }\bibfield  {title} {\bibinfo
		{title} {On-chip interaction-free measurements via the quantum Zeno effect},\
	}\href {https://doi.org/10.1103/PhysRevA.90.042109} {\bibfield  {journal}
		{\bibinfo  {journal} {Phys. Rev. A}\ }\textbf {\bibinfo {volume} {90}},\
		\bibinfo {pages} {042109} (\bibinfo {year} {2014})}\BibitemShut {NoStop}%
\bibitem [{\citenamefont {Peise}\ \emph {et~al.}(2015)\citenamefont {Peise},
  \citenamefont {L{\"u}cke}, \citenamefont {Pezz{\'e}}, \citenamefont
  {Deuretzbacher}, \citenamefont {Ertmer}, \citenamefont {Arlt}, \citenamefont
  {Smerzi}, \citenamefont {Santos},\ and\ \citenamefont {Klempt}}]{Peise2015}%
  \BibitemOpen
  \bibfield  {author} {\bibinfo {author} {\bibfnamefont {J.}~\bibnamefont
  {Peise}}, \bibinfo {author} {\bibfnamefont {B.}~\bibnamefont {L{\"u}cke}},
  \bibinfo {author} {\bibfnamefont {L.}~\bibnamefont {Pezz{\'e}}}, \bibinfo
  {author} {\bibfnamefont {F.}~\bibnamefont {Deuretzbacher}}, \bibinfo {author}
  {\bibfnamefont {W.}~\bibnamefont {Ertmer}}, \bibinfo {author} {\bibfnamefont
  {J.}~\bibnamefont {Arlt}}, \bibinfo {author} {\bibfnamefont {A.}~\bibnamefont
  {Smerzi}}, \bibinfo {author} {\bibfnamefont {L.}~\bibnamefont {Santos}},
  and \bibinfo {author} {\bibfnamefont {C.}~\bibnamefont {Klempt}},\
  }\bibfield  {title} {\bibinfo {title} {Interaction-free
  measurements by quantum Zeno stabilization of ultracold atoms},\ }\href
  {\doibase 10.1038/ncomms7811} {\bibfield  {journal} {\bibinfo  {journal}
  {Nature Communications}\ }\textbf {\bibinfo {volume} {6}},\ \bibinfo {pages}
  {6811} (\bibinfo {year} {2015})}\BibitemShut {NoStop}%
\bibitem [{\citenamefont {Hardy}(1992)}]{Hardy_1992}%
  \BibitemOpen
  \bibfield  {author} {\bibinfo {author} {\bibfnamefont {L.}\ \bibnamefont
  {Hardy}},\ }\bibfield  {title} {\bibinfo {title} {Quantum
  mechanics, local realistic theories, and Lorentz-invariant realistic
  theories},\ }\href {\doibase 10.1103/PhysRevLett.68.2981} {\bibfield
  {journal} {\bibinfo  {journal} {Phys. Rev. Lett.}\ }\textbf {\bibinfo
  {volume} {68}},\ \bibinfo {pages} {2981--2984} (\bibinfo {year}
  {1992})}\BibitemShut {NoStop}%
\bibitem [{\citenamefont {Elouard}\ \emph {et~al.}(2020)\citenamefont
  {Elouard}, \citenamefont {Waegell}, \citenamefont {Huard},\ and\
  \citenamefont {Jordan}}]{Elouard_2020}%
  \BibitemOpen
  \bibfield  {author} {\bibinfo {author} {\bibfnamefont {C.}\ \bibnamefont
  {Elouard}}, \bibinfo {author} {\bibfnamefont {M.}\ \bibnamefont
  {Waegell}}, \bibinfo {author} {\bibfnamefont {B.}\ \bibnamefont
  {Huard}}, and \bibinfo {author} {\bibfnamefont {A.~N.}\ \bibnamefont
  {Jordan}},\ }\bibfield  {title} {\bibinfo {title} {An
  interaction-free quantum measurement-driven engine},\ }\href {\doibase
  10.1007/s10701-020-00381-1} {\bibfield  {journal} {\bibinfo  {journal}
  {Foundations of Physics}\ }\textbf {\bibinfo {volume} {50}},\ \bibinfo
  {pages} {1294--1314} (\bibinfo {year} {2020})}\BibitemShut {NoStop}%
\bibitem [{\citenamefont {Aharonov}\ \emph {et~al.}(2018)\citenamefont
  {Aharonov}, \citenamefont {Cohen}, \citenamefont {Elitzur},\ and\
  \citenamefont {Smolin}}]{Aharonov_2018}%
  \BibitemOpen
  \bibfield  {author} {\bibinfo {author} {\bibfnamefont {Y.}\ \bibnamefont
  {Aharonov}}, \bibinfo {author} {\bibfnamefont {E.}\ \bibnamefont
  {Cohen}}, \bibinfo {author} {\bibfnamefont {A.~C.}\ \bibnamefont
  {Elitzur}}, and \bibinfo {author} {\bibfnamefont {Lee}\ \bibnamefont
  {Smolin}},\ }\bibfield  {title} {\bibinfo {title} {Interaction-free
  effects between distant atoms},\ }\href {\doibase 10.1007/s10701-017-0127-y}
  {\bibfield  {journal} {\bibinfo  {journal} {Foundations of Physics}\ }\textbf
  {\bibinfo {volume} {48}},\ \bibinfo {pages} {1--16} (\bibinfo {year}
  {2018})}\BibitemShut {NoStop}%
\bibitem [{\citenamefont {Blumenthal}\ \emph {et~al.}(2022)\citenamefont
  {Blumenthal}, \citenamefont {Mor}, \citenamefont {Diringer}, \citenamefont
  {Martin}, \citenamefont {Lewalle}, \citenamefont {Burgarth}, \citenamefont
  {Whaley},\ and\ \citenamefont {Hacohen-Gourgy}}]{Blumenthal_2022}%
  \BibitemOpen
  \bibfield  {author} {\bibinfo {author} {\bibfnamefont {E.}~\bibnamefont
  {Blumenthal}}, \bibinfo {author} {\bibfnamefont {C.}~\bibnamefont {Mor}},
  \bibinfo {author} {\bibfnamefont {A.~A.}\ \bibnamefont {Diringer}}, \bibinfo
  {author} {\bibfnamefont {L.~S.}\ \bibnamefont {Martin}}, \bibinfo {author}
  {\bibfnamefont {P.}~\bibnamefont {Lewalle}}, \bibinfo {author} {\bibfnamefont
  {D.}~\bibnamefont {Burgarth}}, \bibinfo {author} {\bibfnamefont {K.~B.}\
  \bibnamefont {Whaley}}, and \bibinfo {author} {\bibfnamefont
  {S.}~\bibnamefont {Hacohen-Gourgy}},\ }\bibfield  {title} {\bibinfo
  {title} {Demonstration of universal control between non-interacting qubits
  using the quantum Zeno effect},\ }\href {\doibase
  10.1038/s41534-022-00594-4} {\bibfield  {journal} {\bibinfo  {journal} {npj
  Quantum Information}\ }\textbf {\bibinfo {volume} {8}},\ \bibinfo {pages}
  {88} (\bibinfo {year} {2022})}\BibitemShut {NoStop}%
\bibitem [{\citenamefont {Burgarth}\ \emph {et~al.}(2014)\citenamefont
  {Burgarth}, \citenamefont {Facchi}, \citenamefont {Giovannetti},
  \citenamefont {Nakazato}, \citenamefont {Pascazio},\ and\ \citenamefont
  {Yuasa}}]{Burgarth_2014}%
  \BibitemOpen
  \bibfield  {author} {\bibinfo {author} {\bibfnamefont {D.~K.}\
  \bibnamefont {Burgarth}}, \bibinfo {author} {\bibfnamefont {P.}\
  \bibnamefont {Facchi}}, \bibinfo {author} {\bibfnamefont {V.}\
  \bibnamefont {Giovannetti}}, \bibinfo {author} {\bibfnamefont {H.}\
  \bibnamefont {Nakazato}}, \bibinfo {author} {\bibfnamefont {S.}\
  \bibnamefont {Pascazio}}, and \bibinfo {author} {\bibfnamefont {K.}\
  \bibnamefont {Yuasa}},\ }\bibfield  {title} {\bibinfo {title}
  {Exponential rise of dynamical complexity in quantum computing through
  projections},\ }\href {\doibase 10.1038/ncomms6173} {\bibfield  {journal}
  {\bibinfo  {journal} {Nature Communications}\ }\textbf {\bibinfo {volume}
  {5}},\ \bibinfo {pages} {5173} (\bibinfo {year} {2014})}\BibitemShut
  {NoStop}%
\bibitem [{\citenamefont {White}\ \emph {et~al.}(1998)\citenamefont {White},
  \citenamefont {Mitchell}, \citenamefont {Nairz},\ and\ \citenamefont
  {Kwiat}}]{White1998}%
  \BibitemOpen
  \bibfield  {author} {\bibinfo {author} {\bibfnamefont {A.~G.}\
  \bibnamefont {White}}, \bibinfo {author} {\bibfnamefont {J.~R.}\
  \bibnamefont {Mitchell}}, \bibinfo {author} {\bibfnamefont {O.}\
  \bibnamefont {Nairz}}, \ and\ \bibinfo {author} {\bibfnamefont {P.~G.}\
  \bibnamefont {Kwiat}},\ }\bibfield  {title} {\bibinfo {title}
  {``Interaction-free'' imaging},\ }\href {\doibase 10.1103/PhysRevA.58.605}
  {\bibfield  {journal} {\bibinfo  {journal} {Phys. Rev. A}\ }\textbf {\bibinfo
  {volume} {58}},\ \bibinfo {pages} {605--613} (\bibinfo {year}
  {1998})}\BibitemShut {NoStop}%
\bibitem [{\citenamefont {Noh}(2009)}]{Noh2009}%
  \BibitemOpen
  \bibfield  {author} {\bibinfo {author} {\bibfnamefont {T.-G.}\ \bibnamefont
  {Noh}},\ }\bibfield  {title} {\bibinfo {title} {Counterfactual
  quantum cryptography},\ }\href {\doibase 10.1103/PhysRevLett.103.230501}
  {\bibfield  {journal} {\bibinfo  {journal} {Phys. Rev. Lett.}\ }\textbf
  {\bibinfo {volume} {103}},\ \bibinfo {pages} {230501} (\bibinfo {year}
  {2009})}\BibitemShut {NoStop}%
\bibitem [{\citenamefont {Li}\ \emph {et~al.}(2020)\citenamefont {Li},
  \citenamefont {Wang}, \citenamefont {Xu}, \citenamefont {Yang}, \citenamefont
  {Al-Amri},\ and\ \citenamefont {Zubairy}}]{zheng-pra-2020}%
  \BibitemOpen
  \bibfield  {author} {\bibinfo {author} {\bibfnamefont {Z.-H.}\
  \bibnamefont {Li}}, \bibinfo {author} {\bibfnamefont {L.}\ \bibnamefont
  {Wang}}, \bibinfo {author} {\bibfnamefont {J.}\ \bibnamefont {Xu}},
  \bibinfo {author} {\bibfnamefont {Y.}\ \bibnamefont {Yang}}, \bibinfo
  {author} {\bibfnamefont {M.}~\bibnamefont {Al-Amri}}, \ and\ \bibinfo
  {author} {\bibfnamefont {M.~S.}\ \bibnamefont {Zubairy}},\ }\bibfield
  {title} {\bibinfo {title} {Counterfactual Trojan horse attack},\
  }\href {\doibase 10.1103/PhysRevA.101.022336} {\bibfield  {journal} {\bibinfo
   {journal} {Phys. Rev. A}\ }\textbf {\bibinfo {volume} {101}},\ \bibinfo
  {pages} {022336} (\bibinfo {year} {2020})}\BibitemShut {NoStop}%
\bibitem [{\citenamefont {Zhang}\ \emph {et~al.}(2019)\citenamefont {Zhang},
  \citenamefont {Sit}, \citenamefont {Bouchard}, \citenamefont {Larocque},
  \citenamefont {Grenapin}, \citenamefont {Cohen}, \citenamefont {Elitzur},
  \citenamefont {Harden}, \citenamefont {Boyd},\ and\ \citenamefont
  {Karimi}}]{Zhang_2019}%
  \BibitemOpen
  \bibfield  {author} {\bibinfo {author} {\bibfnamefont {Y.}\ \bibnamefont
  {Zhang}}, \bibinfo {author} {\bibfnamefont {A.}\ \bibnamefont {Sit}},
  \bibinfo {author} {\bibfnamefont {F.}\ \bibnamefont
  {Bouchard}}, \bibinfo {author} {\bibfnamefont {H.}\ \bibnamefont
  {Larocque}}, \bibinfo {author} {\bibfnamefont {F.}\ \bibnamefont
  {Grenapin}}, \bibinfo {author} {\bibfnamefont {E.}\ \bibnamefont
  {Cohen}}, \bibinfo {author} {\bibfnamefont {A.~C.}\ \bibnamefont
  {Elitzur}}, \bibinfo {author} {\bibfnamefont {J.~L.}\ \bibnamefont
  {Harden}}, \bibinfo {author} {\bibfnamefont {R.~W.}\ \bibnamefont
  {Boyd}}, \ and\ \bibinfo {author} {\bibfnamefont {E.}\ \bibnamefont
  {Karimi}},\ }\bibfield  {title} {\bibinfo {title} {Interaction-free
  ghost-imaging of structured objects},\ }\href {\doibase
  10.1364/OE.27.002212} {\bibfield  {journal} {\bibinfo  {journal} {Opt.
  Express}\ }\textbf {\bibinfo {volume} {27}},\ \bibinfo {pages} {2212--2224}
  (\bibinfo {year} {2019})}\BibitemShut {NoStop}%
\bibitem [{\citenamefont {Hance}\ and\ \citenamefont
  {Rarity}(2021)}]{hans-npj-2021}%
  \BibitemOpen
  \bibfield  {author} {\bibinfo {author} {\bibfnamefont {J.~R.}\
  \bibnamefont {Hance}}\ and\ \bibinfo {author} {\bibfnamefont {J.}\
  \bibnamefont {Rarity}},\ }\bibfield  {title} {\bibinfo {title}
  {Counterfactual ghost imaging},\ }\href {\doibase
  10.1038/s41534-021-00411-4} {\bibfield  {journal} {\bibinfo  {journal} {npj
  Quantum Information}\ }\textbf {\bibinfo {volume} {7}},\ \bibinfo {pages}
  {88} (\bibinfo {year} {2021})}\BibitemShut {NoStop}%
\bibitem [{\citenamefont {Yang}\ \emph {et~al.}(2023)\citenamefont {Yang},
  \citenamefont {Liang}, \citenamefont {Xu}, \citenamefont {Zhang},
  \citenamefont {Zhu},\ and\ \citenamefont {Ma}}]{Ma2023}%
  \BibitemOpen
  \bibfield  {author} {\bibinfo {author} {\bibfnamefont {Y.}\ \bibnamefont
  {Yang}}, \bibinfo {author} {\bibfnamefont {H.}\ \bibnamefont {Liang}},
  \bibinfo {author} {\bibfnamefont {X.}\ \bibnamefont {Xu}}, \bibinfo
  {author} {\bibfnamefont {L.}\ \bibnamefont {Zhang}}, \bibinfo {author}
  {\bibfnamefont {S.}\ \bibnamefont {Zhu}}, and \bibinfo {author}
  {\bibfnamefont {X.-s.}\ \bibnamefont {Ma}},\ }\bibfield  {title}
  {\bibinfo {title} {Interaction-free, single-pixel quantum imaging
  with undetected photons},\ }\href {\doibase 10.1038/s41534-022-00673-6}
  {\bibfield  {journal} {\bibinfo  {journal} {npj Quantum Information}\
  }\textbf {\bibinfo {volume} {9}},\ \bibinfo {pages} {2} (\bibinfo {year}
  {2023})}\BibitemShut {NoStop}%
\bibitem [{\citenamefont {Salih}\ \emph {et~al.}(2013)\citenamefont {Salih},
  \citenamefont {Li}, \citenamefont {Al-Amri},\ and\ \citenamefont
  {Zubairy}}]{salih-prl-2013}%
  \BibitemOpen
  \bibfield  {author} {\bibinfo {author} {\bibfnamefont {H.}\ \bibnamefont
  {Salih}}, \bibinfo {author} {\bibfnamefont {Z.-H.}\ \bibnamefont {Li}},
  \bibinfo {author} {\bibfnamefont {M.}~\bibnamefont {Al-Amri}}, \ and\
  \bibinfo {author} {\bibfnamefont {M.~S.}\ \bibnamefont {Zubairy}},\
  }\bibfield  {title} {\bibinfo {title} {Protocol for direct
  counterfactual quantum communication},\ }\href {\doibase
  10.1103/PhysRevLett.110.170502} {\bibfield  {journal} {\bibinfo  {journal}
  {Phys. Rev. Lett.}\ }\textbf {\bibinfo {volume} {110}},\ \bibinfo {pages}
  {170502} (\bibinfo {year} {2013})}\BibitemShut {NoStop}%
\bibitem [{\citenamefont {Vaidman}(2015)}]{Vaidman_2015}%
  \BibitemOpen
  \bibfield  {author} {\bibinfo {author} {\bibfnamefont {L.}~\bibnamefont
  {Vaidman}},\ }\bibfield  {title} {\bibinfo {title}
  {Counterfactuality of `counterfactual' communication},\ }\href {\doibase
  10.1088/1751-8113/48/46/465303} {\bibfield  {journal} {\bibinfo  {journal}
  {Journal of Physics A: Mathematical and Theoretical}\ }\textbf {\bibinfo
  {volume} {48}},\ \bibinfo {pages} {465303} (\bibinfo {year}
  {2015})}\BibitemShut {NoStop}%
\bibitem [{\citenamefont {Cao}\ \emph {et~al.}(2017)\citenamefont {Cao},
  \citenamefont {Li}, \citenamefont {Cao}, \citenamefont {Yin}, \citenamefont
  {Chen}, \citenamefont {Yin}, \citenamefont {Chen}, \citenamefont {Ma},
  \citenamefont {Peng},\ and\ \citenamefont {Pan}}]{Cao2017}%
  \BibitemOpen
  \bibfield  {author} {\bibinfo {author} {\bibfnamefont {Y.}\ \bibnamefont
  {Cao}}, \bibinfo {author} {\bibfnamefont {Y.-H.}\ \bibnamefont {Li}},
  \bibinfo {author} {\bibfnamefont {Z.}\ \bibnamefont {Cao}}, \bibinfo
  {author} {\bibfnamefont {J.}\ \bibnamefont {Yin}}, \bibinfo {author}
  {\bibfnamefont {Y.-A.}\ \bibnamefont {Chen}}, \bibinfo {author}
  {\bibfnamefont {H.-L.}\ \bibnamefont {Yin}}, \bibinfo {author}
  {\bibfnamefont {T.-Y.}\ \bibnamefont {Chen}}, \bibinfo {author}
  {\bibfnamefont {X.}\ \bibnamefont {Ma}}, \bibinfo {author}
  {\bibfnamefont {C.-Z.}\ \bibnamefont {Peng}}, \ and\ \bibinfo {author}
  {\bibfnamefont {J.-W.}\ \bibnamefont {Pan}},\ }\bibfield  {title}
  {\bibinfo {title} {Direct counterfactual communication via quantum
  Zeno effect},\ }\href {\doibase 10.1073/pnas.1614560114} {\bibfield
  {journal} {\bibinfo  {journal} {Proceedings of the National Academy of
  Sciences}\ }\textbf {\bibinfo {volume} {114}},\ \bibinfo {pages} {4920--4924}
  (\bibinfo {year} {2017})}\BibitemShut {NoStop}%
\bibitem [{\citenamefont {Aharonov}\ and\ \citenamefont
  {Vaidman}(2019)}]{vaidman-pra-2019}%
  \BibitemOpen
  \bibfield  {author} {\bibinfo {author} {\bibfnamefont {Y.}\ \bibnamefont
  {Aharonov}}\ and\ \bibinfo {author} {\bibfnamefont {L.}\ \bibnamefont
  {Vaidman}},\ }\bibfield  {title} {\bibinfo {title} {Modification of
  counterfactual communication protocols that eliminates weak particle
  traces},\ }\href {\doibase 10.1103/PhysRevA.99.010103} {\bibfield  {journal}
  {\bibinfo  {journal} {Phys. Rev. A}\ }\textbf {\bibinfo {volume} {99}},\
  \bibinfo {pages} {010103} (\bibinfo {year} {2019})}\BibitemShut {NoStop}%
\bibitem [{\citenamefont {Calafell}\ \emph {et~al.}(2019)\citenamefont
  {Calafell}, \citenamefont {Strömberg}, \citenamefont {Arvidsson-Shukur},
  \citenamefont {Rozema}, \citenamefont {Saggio}, \citenamefont {Greganti},
  \citenamefont {Harris}, \citenamefont {Prabhu}, \citenamefont {Carolan},
  \citenamefont {Hochberg}, \citenamefont {Baehr-Jones}, \citenamefont
  {Englund}, \citenamefont {Barnes},\ and\ \citenamefont
  {Walther}}]{Walther2019}%
  \BibitemOpen
  \bibfield  {author} {\bibinfo {author} {\bibfnamefont {I.~A.}\
  \bibnamefont {Calafell}}, \bibinfo {author} {\bibfnamefont {T.}~\bibnamefont
  {Strömberg}}, \bibinfo {author} {\bibfnamefont {D.~R.~M.}\ \bibnamefont
  {Arvidsson-Shukur}}, \bibinfo {author} {\bibfnamefont {L.~A.}\ \bibnamefont
  {Rozema}}, \bibinfo {author} {\bibfnamefont {V.}~\bibnamefont {Saggio}},
  \bibinfo {author} {\bibfnamefont {C.}~\bibnamefont {Greganti}}, \bibinfo
  {author} {\bibfnamefont {N.~C.}\ \bibnamefont {Harris}}, \bibinfo {author}
  {\bibfnamefont {M.}~\bibnamefont {Prabhu}}, \bibinfo {author} {\bibfnamefont
  {J.}~\bibnamefont {Carolan}}, \bibinfo {author} {\bibfnamefont
  {M.}~\bibnamefont {Hochberg}}, \bibinfo {author} {\bibfnamefont
  {\emph {et~al.}}},\ }\bibfield  {title} {\bibinfo
  {title} {Trace-free counterfactual communication with a
  nanophotonic processor},\ }\href {\doibase 10.1038/s41534-019-0179-2}
  {\bibfield  {journal} {\bibinfo  {journal} {npj Quantum Information}\
  }\textbf {\bibinfo {volume} {5}},\ \bibinfo {pages} {61} (\bibinfo {year}
  {2019})}\BibitemShut {NoStop}%
\bibitem [{\citenamefont {Aharonov}\ \emph {et~al.}(2021)\citenamefont
  {Aharonov}, \citenamefont {Cohen},\ and\ \citenamefont
  {Popescu}}]{Aharonov2021}%
  \BibitemOpen
  \bibfield  {author} {\bibinfo {author} {\bibfnamefont {Y.}\ \bibnamefont
  {Aharonov}}, \bibinfo {author} {\bibfnamefont {E.}\ \bibnamefont
  {Cohen}}, \ and\ \bibinfo {author} {\bibfnamefont {S.}\ \bibnamefont
  {Popescu}},\ }\bibfield  {title} {\bibinfo {title} {A dynamical
  quantum Cheshire Cat effect and implications for counterfactual
  communication},\ }\href {\doibase 10.1038/s41467-021-24933-9} {\bibfield
  {journal} {\bibinfo  {journal} {Nature Communications}\ }\textbf {\bibinfo
  {volume} {12}},\ \bibinfo {pages} {4770} (\bibinfo {year}
  {2021})}\BibitemShut {NoStop}%
\bibitem [{\citenamefont {Hosten}\ \emph {et~al.}(2006)\citenamefont {Hosten},
  \citenamefont {Rakher}, \citenamefont {Barreiro}, \citenamefont {A.},\ and\
  \citenamefont {Kwiat}}]{Hosten2006}%
  \BibitemOpen
  \bibfield  {author} {\bibinfo {author} {\bibfnamefont {O.}~\bibnamefont
  {Hosten}}, \bibinfo {author} {\bibfnamefont {M.~T.}\ \bibnamefont {Rakher}},
  \bibinfo {author} {\bibfnamefont {J.~T.}\ \bibnamefont {Barreiro}}, \bibinfo
  {author} {\bibfnamefont {Peters~N.}\ \bibnamefont {A.}}, \ and\ \bibinfo
  {author} {\bibfnamefont {P.~G.}\ \bibnamefont {Kwiat}},\ }\bibfield  {title}
  {\bibinfo {title} {Counterfactual quantum computation through
  quantum interrogation},\ }\href {\doibase 10.1038/nature04523} {\bibfield
  {journal} {\bibinfo  {journal} {Nature}\ }\textbf {\bibinfo {volume} {439}},\
  \bibinfo {pages} {949} (\bibinfo {year} {2006})}\BibitemShut {NoStop}%
\bibitem [{\citenamefont {Dogra}\ \emph {et~al.}(2022)\citenamefont {Dogra},
  \citenamefont {McCord},\ and\ \citenamefont {Paraoanu}}]{our_protocol}%
  \BibitemOpen
  \bibfield  {author} {\bibinfo {author} {\bibfnamefont {S.}\ \bibnamefont
  {Dogra}}, \bibinfo {author} {\bibfnamefont {J.~J.}\ \bibnamefont {McCord}},
  \ and\ \bibinfo {author} {\bibfnamefont {G.~S.}\ \bibnamefont
  {Paraoanu}},\ }\bibfield  {title} {\bibinfo {title} {Coherent
  interaction-free detection of microwave pulses with a superconducting
  circuit},\ }\href {\doibase 10.1038/s41467-022-35049-z} {\bibfield
  {journal} {\bibinfo  {journal} {Nature Communications}\ }\textbf {\bibinfo
  {volume} {13}},\ \bibinfo {pages} {7528} (\bibinfo {year}
  {2022})}\BibitemShut {NoStop}%
\bibitem [{\citenamefont {Wigner}(1963)}]{Wigner_1963}%
  \BibitemOpen
  \bibfield  {author} {\bibinfo {author} {\bibfnamefont {E.~P.}\
  \bibnamefont {Wigner}},\ }\bibfield  {title} {\bibinfo {title} {The
  problem of measurement},\ }\href {\doibase 10.1119/1.1969254} {\bibfield
  {journal} {\bibinfo  {journal} {American Journal of Physics}\ }\textbf
  {\bibinfo {volume} {31}},\ \bibinfo {pages} {6--15} (\bibinfo {year}
  {1963})}\BibitemShut {NoStop}%
\bibitem [{\citenamefont {Paraoanu}(2006)}]{Paraoanu_2006}%
  \BibitemOpen
  \bibfield  {author} {\bibinfo {author} {\bibfnamefont {G.~S.}\ \bibnamefont
  {Paraoanu}},\ }\bibfield  {title} {\bibinfo {title}
  {Interaction-free measurements with superconducting qubits},\ }\href
  {\doibase 10.1103/PhysRevLett.97.180406} {\bibfield  {journal} {\bibinfo
  {journal} {Phys. Rev. Lett.}\ }\textbf {\bibinfo {volume} {97}},\ \bibinfo
  {pages} {180406} (\bibinfo {year} {2006})}\BibitemShut {NoStop}%
\bibitem [{\citenamefont {Wander}\ \emph {et~al.}(2021)\citenamefont {Wander},
  \citenamefont {Cohen},\ and\ \citenamefont {Vaidman}}]{Wander}%
  \BibitemOpen
  \bibfield  {author} {\bibinfo {author} {\bibfnamefont {A.}\ \bibnamefont
  {Wander}}, \bibinfo {author} {\bibfnamefont {E.}\ \bibnamefont {Cohen}},
  and \bibinfo {author} {\bibfnamefont {L.}\ \bibnamefont {Vaidman}},\
  }\bibfield  {title} {\bibinfo {title} {Three approaches for
  analyzing the counterfactuality of counterfactual protocols},\ }\href
  {\doibase 10.1103/PhysRevA.104.012610} {\bibfield  {journal} {\bibinfo
  {journal} {Phys. Rev. A}\ }\textbf {\bibinfo {volume} {104}},\ \bibinfo
  {pages} {012610} (\bibinfo {year} {2021})}\BibitemShut {NoStop}%
\bibitem [{\citenamefont {Sultanov}\ \emph {et~al.}(2021)\citenamefont
  {Sultanov}, \citenamefont {Kuzmanović}, \citenamefont {Lebedev},\ and\
  \citenamefont {Paraoanu}}]{Sultanov2021}%
  \BibitemOpen
  \bibfield  {author} {\bibinfo {author} {\bibfnamefont {A.}\ \bibnamefont
  {Sultanov}}, \bibinfo {author} {\bibfnamefont {M.}\ \bibnamefont
  {Kuzmanović}}, \bibinfo {author} {\bibfnamefont {A.~V.}\ \bibnamefont
  {Lebedev}}, and \bibinfo {author} {\bibfnamefont {G.~S.}\
  \bibnamefont {Paraoanu}},\ }\bibfield  {title} {\bibinfo {title}
  {Protocol for temperature sensing using a three-level transmon circuit},\
  }\href {\doibase 10.1063/5.0065224} {\bibfield  {journal} {\bibinfo
  {journal} {Applied Physics Letters}\ }\textbf {\bibinfo {volume} {119}},\
  \bibinfo {pages} {144002} (\bibinfo {year} {2021})}\BibitemShut {NoStop}%
\bibitem [{\citenamefont {Li}\ \emph {et~al.}(2012)\citenamefont {Li},
  \citenamefont {Sillanpää}, \citenamefont {Paraoanu},\ and\ \citenamefont
  {Hakonen}}]{Li2012}%
  \BibitemOpen
  \bibfield  {author} {\bibinfo {author} {\bibfnamefont {J.}\ \bibnamefont
  {Li}}, \bibinfo {author} {\bibfnamefont {M.~A.}\ \bibnamefont
  {Sillanpää}}, \bibinfo {author} {\bibfnamefont {G.~S.}\ \bibnamefont
  {Paraoanu}}, and \bibinfo {author} {\bibfnamefont {P.~J.}\ \bibnamefont
  {Hakonen}},\ }\bibfield  {title} {\bibinfo {title} {Pure dephasing
  in a superconducting three-level system},\ }\href {\doibase
  10.1088/1742-6596/400/4/042039} {\bibfield  {journal} {\bibinfo  {journal}
  {Journal of Physics: Conference Series}\ }\textbf {\bibinfo {volume} {400}},\
  \bibinfo {pages} {042039} (\bibinfo {year} {2012})}\BibitemShut {NoStop}%
\bibitem [{\citenamefont {Tempel}\ and\ \citenamefont
  {Aspuru-Guzik}(2011)}]{Guzik2011}%
  \BibitemOpen
  \bibfield  {author} {\bibinfo {author} {\bibfnamefont {D.~G.}\
  \bibnamefont {Tempel}} and \bibinfo {author} {\bibfnamefont {A.}\
  \bibnamefont {Aspuru-Guzik}},\ }\bibfield  {title} {\bibinfo
  {title} {Relaxation and dephasing in open quantum systems time-dependent
  density functional theory: Properties of exact functionals from an
  exactly-solvable model system},\ }\href {\doibase
  https://doi.org/10.1016/j.chemphys.2011.03.014} {\bibfield  {journal}
  {\bibinfo  {journal} {Chemical Physics}\ }\textbf {\bibinfo {volume} {391}},\
  \bibinfo {pages} {130--142} (\bibinfo {year} {2011})}\BibitemShut {NoStop}%
\end{thebibliography}
\end{document}